\documentclass[traditabstract]{aa}
\usepackage{epsfig}
\usepackage{graphicx,alltt}
\pdfoutput=1

\usepackage{graphicx,natbib}
\usepackage{natbib} 
\usepackage{amsmath,amssymb}

\def\beq{\begin{equation}}
\def\eeq{\end{equation}}
\def\simgr{\,\hbox{\hbox{$ > $}\kern -0.8em \lower 1.0ex\hbox{$\sim$}}\,}
\def\simle{\,\hbox{\hbox{$ < $}\kern -0.8em \lower 1.0ex\hbox{$\sim$}}\,}

\titlerunning{The contribution of galaxies escaping UV/optical color selections at $z\sim2$}
\authorrunning{Riguccini et al.}

\begin{document}

\title{Dust-obscured star formation and the contribution of galaxies escaping UV/optical color selections at $z\sim2$}

\author{L. Riguccini\inst{1}, E. Le Floc'h\inst{1}, O. Ilbert\inst{2}, H. Aussel\inst{1}, M. Salvato\inst{3}, P. Capak\inst{4}, H. McCracken\inst{5}, J. Kartaltepe\inst{6}, D. Sanders\inst{7}, N. Scoville\inst{4}
}

\institute{Laboratoire AIM, CEA/DSM-CNRS-Universit\'e Paris Diderot, IRFU/Service d'Astrophysique, B\^at.709, CEA-Saclay, 91191 Gif-sur-Yvette Cedex, France (laurie.riguccini, emeric.lefloch@cea.fr)
\and Laboratoire d'Astrophysique de Marseille, BP 8, Traverse du Siphon, 13376 Marseille Cedex 12, France
\and Max-Planck-Institute f\"ur Plasma Physics, Boltzmann Strasse 2, Garching 85748, Germany
\and California Institute of Technology, MC 105-24, 1200 East California Boulevard, Pasadena, CA 91125, USA
\and  Institut d'Astrophysique de Paris, UMR7095 CNRS, Universit\'e Pierre et Marie Curie, 98 bis Boulevard Arago, 75014 Paris, France
\and  National Optical Astronomy Observatory, 950 N. Cherry Ave., Tucson, AZ, 85719, USA
\and Institute for Astronomy, 2680 Woodlawn Dr., University of Hawaii, Honolulu, HI 96822, USA
}

\date{Accepted for publication in Astronomy \& Astrophysics, May 31, 2011}

\offprints{L. Riguccini}

\abstract{
{\it Context.} A substantial fraction of the stellar mass growth across cosmic time occurred within dust-enshrouded environments. So far the  identification of complete samples of distant star-forming galaxies from the short wavelength range has thus been  strongly biased by the effect of dust extinction.  Yet, the exact amount of star-forming activity that took place in high-redshift dusty galaxies currently missed by optical surveys has been barely explored.

{\it Aims.} Our goal is to determine the fraction of luminous star-forming galaxies at $1.5 \lesssim z \lesssim 3$  potentially missed by the traditional color selection techniques because of dust extinction. We also aim  at quantifying the contribution of these sources to the IR luminosity and cosmic star formation density at high redshift.

{\it Methods.} We base our work on a sample of 24$\mu$m sources brighter than 80$\mu$Jy and taken from the {\it Spitzer} survey of the COSMOS field. Almost all of these sources have accurate photometric redshifts.  We apply to this mid-IR selected sample the BzK and  BM/BX criteria as well as the selections of the ``{\it IRAC peakers\,}" and the ``{\it Optically-Faint IR-bright\,}" galaxies, and we analyze the fraction of sources identified with these techniques.
We also computed 8$\mu$m rest-frame luminosity from the 24$\mu$m fluxes of our sources, and considering the relationships between $L_{8\mu m}$ and $L_{Pa_{\alpha}}$ and between $L_{8\mu m}$ and $L_{IR}$ we derived $\rho_{IR}$ and then $\rho_{SFR}$ for our MIPS sources. 

{\it Results.} The BzK criterion offers an almost complete ($\sim$90\%) identification of the 24$\mu$m sources at $1.4<z<2.5$. On the contrary, the BM/BX criterion miss 50\% of the MIPS sources. We attribute this bias to the effect of extinction which redden the typical colors of galaxies. The contribution of these two selections to the IR luminosity density produced by all the sources brighter than 80$\mu$Jy are from the same order. Moreover the criterion based on the presence of a stellar bump in their spectra (``{\it IRAC peakers\,}" ) miss up to 40\% of the IR luminosity density while only 25\% of the IR luminosity density at z$\sim$2 is produced by ``{\it Optically-Faint IR-bright\,}" galaxies characterized by extreme mid-IR to optical flux ratios.

{\it Conclusions.} Color selections of distant star-forming galaxies must be used with lots of care given the substantial bias  they can suffer. 
In particular, the effect of dust extinction strongly impacts the completeness of identifications at the bright end of the bolometric luminosity function, which implies large and uncertain extrapolations to account for the contribution of  dusty galaxies missed by these selections. In the context of forthcoming facilities that will operate at long wavelengths (e.g., $JWST$, ALMA, SAFARI, EVLA, SKA), this emphasizes the importance of minimizing the extinction biases when probing the activity of star formation in the early Universe.
}

\keywords{Galaxies: high-redshift - Infrared: galaxies - Cosmology: observations}

\maketitle

\section{Introduction}
\label{sec:intro}

It is well established that  the peak of star-forming activity in the Universe and the bulk of stellar mass assembly in galaxies occurred at $1<z<3$ \citep[e.g.,][]{Madau:96,Steidel:99,Dickinson:03,Hopkins:06,Arnouts:2007}. However the role and the contribution of the different processes that governed this build-up of stellar mass are still open questions. Improving our physical understanding of the population of star-forming galaxies that contributed to the growth of structures in the distant Universe thus remains a critical issue for modern extragalactic astrophysics.
 
To reach this goal, the broad variety of physical properties observed at high redshift usually implies the use of large and {\it complete\,} samples of galaxies mainly selected as a function of star formation rate (SFR) or stellar mass \citep[see for instance][]{Elbaz:07,Noeske:07,Ilbert:10}. This requires systematic identifications of sources with spectroscopic or photometric redshifts \citep[e.g.,][]{Wolf:03,Fontana:04,Glaze:04,Cimatti:08}.  However, the determination of reliable spectroscopic redshifts for such large samples of distant galaxies is time-consuming because of the difficulty to identify emission or absorption lines with high enough signal to noise. Flux limited optical selections are also dominated by populations of low-to-intermediate redshift sources and they result in statistically small numbers of distant objects, while the limited coverage of spectroscopic surveys makes them sensitive to the cosmic variance effect. Photometric redshifts, on the other hand, can suffer various systematics and contamination from catastrophic failures depending on the wavelength range and the number of bands they are based on, and their uncertainties can become significant in the case of distant galaxies.

To minimize these difficulties, various techniques based on single- or two-color criteria have been proposed to identify specific populations of galaxies in the early Universe. Among them the first and best-known example is probably the selection of Lyman Break Galaxies from the typical $U_{n}GR$ colors of sources at $z \sim 3$ \citep{Steidel:03}. Similarly, other techniques have been implemented to select galaxies at more intermediate redshifts ($1.5 \lesssim z \lesssim 3$) based on optical and infrared (IR) broad-band photometry. They include for instance the $BzK$ criterion \citep{Daddi:2004}, the selection of BM/BX sources \citep{Steidel:04,Adel:04} and {\it Distant Red Galaxies\,} \citep[DRGs,][]{Franx:03},  the {\em Extremely Red Objects\,} \citep[EROs,][]{Thompson:99} or the identification of massive sources through the shape of their stellar bump signature in the rest-frame near-Infrared \citep{Simpson:99,Sawicki:02,Huang:04}.  Each of these methods was designed to specifically isolate sub-populations of high-redshift galaxies based on a well-defined characteristic of their spectral energy distribution (e.g., the blue color of their UV continuum, the red color of their evolved stellar populations, signatures of star formation reddened by dust, ...). In this context they have been quite successfully used over the past 10 years and their high efficiency has dramatically revolutionized our understanding of galaxy formation in the early Universe. 
 
 The main advantage of these methods is their straight forward application which requires only a small number of observing bands. Conversely, their calibration over a restricted wavelength range also implies that  they can hardly account for the global diversity of galaxy properties at high redshift, making perfidious the use of these techniques beyond the scope for which they were originally  defined.  In particular it is known that observations at rest-frame UV/optical wavelengths can be strongly affected by extinction. These techniques could thus suffer from biases due to the effect of dust, especially at the bright end of the bolometric luminosity function of star-forming galaxies where the contribution of dust-enshrouded star formation becomes dominant. Indeed, observations of the deep Universe taken with the {\it Spitzer Space Telescope\,}  revealed the existence of  a high-redshift population of dusty luminous  sources yet almost invisible at shorter wavelengths and thus escaping the traditional  UV/optical selections  \citep{Houck:05,Dey:08}.
What is the fraction of  star formation density produced at high redshift  by these highly obscured sources? To what extent do galaxies currently missed by  the standard color selection techniques contribute to the growth of stellar mass? 

 Over the last decades, observations in the Mid- and Far-Infrared as well as at sub-millimeter wavelengths enabled the identification of a large number of luminous and dusty galaxies\footnote{These sources are referred as Luminous Infrared Galaxies (LIRGs: $10^{11}L_{\odot}<L_{IR}<10^{12}L_{\odot}$ with $L_{IR}=L_{8-1000\mu m}$) and Ultra-Luminous Infrared Galaxies (ULIRGs:  $L_{IR} > 10^{12}L_{\odot}$).} throughout cosmic history and up to very large cosmological distances  \citep[e.g.,][]{Smail:97,Hughes:98,Aussel:99,Chary:2001,Blain:02,LeFloch:04,Marleau:04,Coppin:2008,Capak:11}. Detailed studies of these sources using multi-wavelength photometric and spectroscopic surveys have revealed that the bulk of their infrared luminosity originates from intense episodes of massive star formation, while they can also host powerful active nuclei triggered by nucelar accretion onto their central black holes \citep[e.g.,][]{Yan:05,Pope:08,Menendez:09,Hainline:09,Desai:09,Fadda:10}.   Although they have become rare objects at present day, these LIRGs and ULIRGs were quite numerous in the past history of the Universe, dominating the comoving infrared energy density beyond $z\sim0.5$ and making up to 70$\%$ of star-forming activity at $z\sim1$ \citep{Emeric:05}. At higher redshift their contribution could be even larger \citep{Caputi:07,Rodi:10}, although the lack of sensitivity of mid-IR and far-IR experiments has prevented reaching definitive conclusions on this issue. In fact the respective contributions of  dust-obscured and unobscured star formation to the growth of structures has been a long debate for more than 10~years \citep[e.g.][]{AdelSteidel:00}. Whereas  a general consensus recognizes that dusty and luminous star-forming galaxies played a critical role in driving massive galaxy evolution, it is also clear that at the peak of galaxy formation the faint-end slope of the UV luminosity function was much steeper than observed in the local Universe, implying a larger contribution of faint galaxies to the cosmic star formation density \citep{Reddy:08}.  Unfortunately,          far-IR observations have not enabled   the direct probe of these faint high-redshift galaxies yet, while
the characterization of  luminous star-forming galaxies solely based on UV observations requires large extrapolations due to dust extinction. So far it has been therefore difficult to reach consistent pictures from deep surveys performed at UV and far-IR wavelengths.

The goal of this paper is to address the fraction of luminous high-redshift galaxies that may be missed by the different color selection techniques commonly used to identify distant sources, and to quantify the impact of this bias by estimating the contribution of these objects to the IR luminosity and cosmic star formation density at $1.5 \lesssim z \lesssim 3$. We carried out this work using the deep $24\mu m$ observations ($F_{24\mu m} > 0.08$\,mJy) of the COSMOS field \citep{Sco:07a} obtained with the MIPS instrument \citep{Rieke:04} on-board the {\it Spitzer Space Telescope}. Among all facilities allowing the probe of dusty galaxies in the distant Universe (i.e.  SCUBA, AzTEC, LABOCA, MIPS 70 and 160$\mu$m, ...), MIPS 24$\mu$m observations provide one of the deepest sensitivity limits currently reachable up to $z\sim2-3$. Furthermore the deep optical/near-IR imaging as well as the high-quality photometric redshifts obtained in the COSMOS field \citep{Ilbert:09} enabled a detailed characterization of the redshift distribution associated to the whole sample of 24$\mu$m sources up to $z\sim 3$ \citep{Emeric:09}, hence providing a complete and unbiased view of the population of dusty high-redshift sources selected at $F_{24\mu m} > 0.08$\,mJy.  Taking advantage of the large coverage of COSMOS to minimize the effect of cosmic variance, we applied the color criteria associated to four different selections of distant galaxies ($BzK$ \& BM/BX sources, ``{\it IRAC peakers\,}" and ``{\it Optically-Faint IR-bright\,}" (OFIR) objects), and we quantified the amount of high-redshift 24$\mu$m sources missed by each of these techniques.  Our data are described in Sect.\,\ref{sec:dat}, the criteria that we explored are presented in Sect.\,\ref{sec:select1} and the corresponding sub-selections that we obtained based on the MIPS-detected COSMOS population are discussed in Sect.\,4.  In Sect.\,\ref{sec:LF} we present the rest-frame 8$\mu m$ luminosity function (LF) of the complete high-redshift MIPS sample as well as the contribution of the sub-populations of galaxies respectively identified with the color selection techniques mentioned earlier.  Our results are discussed in Sect.\,\ref{sec:discussion} and we finally present our conclusions in Sect.\,\ref{sec:ccl}.  Throughout this paper we assume a $\Lambda$CDM cosmology with H$_{0}$ = 70 km s$^{-1}$, $\Omega_{m}$ = 0.3 and $\Omega_{\lambda}$ = 0.7. Magnitudes are given in the AB system and the conversions between luminosities and Star Formation Rates are computed assuming the Initial Mass Function from \citet{Salpeter:55}.

\section{The Data}
\label{sec:dat}

The sample of luminous star-forming galaxies used in this work was selected from the deep $Spitzer$/MIPS observations of the 2 deg$^{2}$ COSMOS field \citep{Sanders:07} and it is based on the 24$\mu$m source catalog obtained by \citet{Emeric:09}. COSMOS is the largest contiguous imaging survey ever undertaken with the HST \citep{Sco:07b} to a depth of i$^{+}$=28.0 [AB] \citep[5$\sigma$ on an optimally extracted point source;][]{Capak:07}. The COSMOS field is located near the celestial equator to ensure visibility by all ground and spaced-based astronomical facilities. It is devoid of bright X-ray, UV, and radio foreground sources, and compared to other equatorial fields the Galactic extinction is remarkably low and uniform ($\langle E_{(B-V)} \rangle \simeq 0.02$). Extensive multi-wavelength follow-ups with ground-based and spaced-based facilities have also been achieved with high-sensitivity imaging and spectroscopy spanning the entire spectrum from the X-ray to the radio \citep{Hasinger:07,Schinnerer:07,Lilly:07,Elvis:09}. In particular, deep UV-to-8$\mu$m photometry was obtained over 30 different bands using narrow, medium and broad-band filters  \citep[e.g.,][]{Taniguchi:07,Capak:07}, while far-IR and submillimeter/millimeter observations were performed with various instruments such as MIPS on-board $Spitzer$ \citep{Sanders:07}, PACS and SPIRE on-board $Herschel$ (Lutz et al. submitted, Oliver et al. in prep.), SCUBA-2, AzTEC, MAMBO and BOLOCAM \citep[e.g.,][]{Bertoldi:07,Austermann:09}.

While the 24$\mu$m source catalog of COSMOS covers a total area of  2\,deg$^{2}$,
we restricted our sample to the field outside the regions contaminated by very bright or saturated objects and where the photometry is less accurate.
We also applied a conservative flux cut of 80$\mu$Jy, corresponding to a  completeness of $\sim90\%$ in the source extraction performed by \citet{Emeric:09}. This led to a sample of 29\,395 sources detected at 24$\mu$m over an effective surface of 1.68\,deg$^{2}$.

The counterparts of the MIPS-selected detections were identified following the same procedure as the one described by \citet{Emeric:09}. Given the much higher density of sources detected at optical wavelengths in COSMOS \citep{Capak:07} compared to those detected with MIPS,
a direct cross-correlation between the 24$\mu$m--selected catalog and the optical observations could lead to a large number of  spurious associations with optically-detected galaxies randomly aligned close to the line of sight of the MIPS sources. Hence, we first cross-correlated the 24$\mu$m  data with the COSMOS $K_s$-band catalogue of \citet{McCra:10} 
to minimize the risk of wrong associations. The density of  sources in this catalogue is substantially smaller than at optical wavelengths, while it is deep enough (5$\sigma$ for $K_s$=23.7) to allow the identification of near-IR counterparts for most of the
24$\mu$m  detections in COSMOS \citep[e.g.,][]{Emeric:09}.  Also, the $K_s$--band observations were carried out under very good seeing conditions ($\sim$0.7$\arcsec$ at 2.2$\mu$m), leading to a PSF much narrower than obtained in the other near-IR images of COSMOS (e.g., IRAC-3.6$\mu$m) and allowing more robust identifications in the case of blended sources. Our correlation between the 24$\mu$m and $K_s$--band data was performed with a matching radius of 2$\arcsec$. Given  the width of the 24$\mu$m PSF (FWMH$\sim$6"), this radius allowed us to identify most of the 24$\mu$m sources while also minimizing the number of multiple matches. Similar to the results obtained by \citet{Emeric:09}, only 765 sources from the initial sample of  29\,395 mid-IR objects could not be matched in the near-IR, while up to 84\% sources from the MIPS catalogue were identified with a clean and single K$_s$-band counterpart. For the rest of the sample (13\%), two possible matches were found within the matching distance. For those cases we decided to keep the closest counterpart, as for most of them ($\sim$60\%) the centroid was at least twice closer to the 24$\mu$m source than the second possible match.

In a second step we correlated the list of these $K_s$--band identifications with a recently-updated version of the i$^{+}$--band selected catalogue of photometric redshifts from \citet{Ilbert:09} using a matching radius of 1$\arcsec$. Given the depth of the COSMOS i$^{+}$--band data \citep[e.g.,][]{Capak:07} at the end 1306 MIPS sources could not be matched at optical wavelengths, leading to the identification of optical counterparts and redshift determination for more than 95$\%$ of the initial 24$\mu$m--selected sources.  As part of this second cross-correlation, double optical matches were obtained for only $\sim$1\% of the K$_s$-band counterparts, underlying the robustness of our identification at these short wavelengths. For these few cases we kept again the closest possible optical association. The catalog of \citet{Ilbert:09} includes the band-merged COSMOS photometry published by \citet{Capak:07}, \citet{McCra:10} and \citet{Ilbert:10} from the $U$-band to the IRAC 8$\mu$m. Our cross-correlation thus yielded direct determination of optical/near-IR broad-band magnitudes for all selected galaxies. In particular, we considered the photometry in the $u^{*}$, $B_{J}$, $V_{J}$, $r^{+}$, $i^{+}$ and $z^{+}$ bands\footnote{ In this work we also applied to the COSMOS broad-band photometry the systematic offsets inferred by \citet{Ilbert:09} when computing their photometric redshifts (see their Table\,1).}, where the 5-$\sigma$ magnitude limit reaches respectively 26.5, 26.6, 26.5, 26.6, 26.1 and 25.1 \citep{Capak:07}.  Finally, \citet{Ilbert:09} also provide the identification of sources detected with XMM over COSMOS \citep[e.g.,][]{Brusa:10,Brusa:07,Salvato:09}, which allowed us to systematically exclude the X-ray detected AGNs from our final sample down to a flux limited $S_{0.5-2 keV} = 5\times10^{-16}$ erg cm$^{-2}$ s$^{-1}$ (less than 4\%).

\section{Photometric selection techniques}
\label{sec:select1}

\subsection{Optical and NIR color selections}

\subsubsection{The BzK selection}
Based on deep photometry obtained with $B$--band, $z$--band and $K$--band filters, \citet{Daddi:2004} proposed a criterion to select star-forming galaxies in the redshift range of $1.4<z<2.5$: $BzK \equiv (z-K)_{AB} - (B-z)_{AB} > -0.2$. In the plane defined by $z-K$ as a function of $B-z$ (hereafter the $BzK$ diagram), the extinction vector obtained for the attenuation law of \citet{Cal:00} is parallel to the line characterized by $BzK = -0.2$ at $1.5<z<2.5$. Therefore this criterion presents the strength of being mostly independent of dust obscuration.

The $B$--band, $z$--band and $K$--band photometry used by \citet{Daddi:2004} was based on the Bessel--$B$ and $K_{s}$ filters available at the VLT as well as on the F850LP $z$--band filter of HST.  On the other hand, our COSMOS photometry was determined through $B_{J}$ and $z^{+}$--band observations obtained at the Subaru Telescope and $K_{s}$--band data taken at CFHT \citep{Capak:07,McCra:10}.  In order to adapt the original $BzK$ criterion to the COSMOS data, we thus determined the evolution of the $BzK$ variable through the COSMOS filters as a function of redshift and with a set of star-forming galaxy templates similar to the library of SEDs used by \citet{Ilbert:09} for their catalog of photometric redshifts in COSMOS, and we compared this $BzK$ variable to the $BzK_o$ color initially defined by \citet{Daddi:2004}.  The star-forming SEDs used by \citet{Ilbert:09} were generated with the stellar population synthesis model of \citet{Bruzual:2003} assuming exponentially-declining star formation histories and starburst ages between 0.03 and 3 Gyr.  In the redshift range of $1.4<z<2.5$ we found $BzK - BzK_o = 0.1$\,mag with a very small dispersion driven by the choice of SED template. To apply the $BzK$ selection technique to the COSMOS survey we thus added a systematic offset of +0.1\,mag to the initial criterion proposed by \citet{Daddi:2004}, leading 
to the identification of star-forming $BzK$ galaxies with the following color:

\begin{equation}
BzK \geqslant -0.1
\end{equation}

\subsubsection{The BM/BX criteria}

Another widely-used selection of distant star-forming sources was proposed by \citet{Steidel:04} and \citet{Adel:04} using UV/optical broad-band color selection techniques. Based on photometry obtained in the $U_{n}$, $G$ and $R$ bands and using the associated $U_n-G$ and $G-R$ colors, they defined two criteria refered as the ``BM'' and ``BX'' selections to identify galaxies at 1.4 $\le$ z $\le$ 2.1 and 1.9 $\le$ z $\le$ 2.7 respectively.

\begin{figure*}
   \centering
   \includegraphics[width=0.8\textwidth]{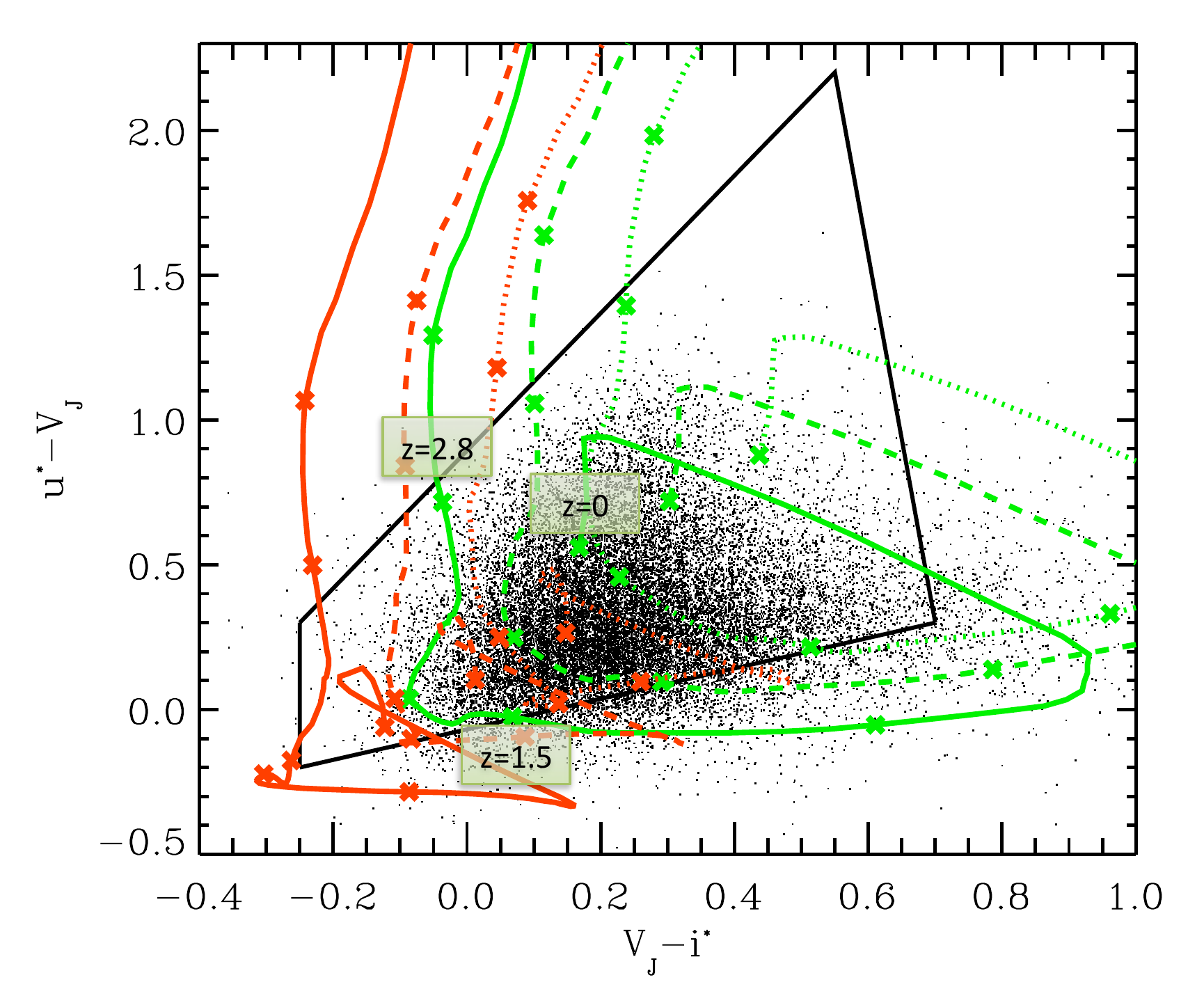} 
   \caption{Selection of galaxies at $1.5<z<2.8$  based on $u^{*} - V_{J}$ and $V_{J} - i^{+}$ colors. The galaxy tracks represented by the green and red solid curves correspond to the colors predicted  as a function of redshift for two  typical  star-forming galaxy templates used by \citet{Ilbert:09} for their determination of photometric redshifts in COSMOS. The dashed and dotted lines show the effect of dust  extinction produced by respectively $E(B-V)$\,=\,0.1 and 0.2 assuming the law of \citet{Cal:00}. The cross symbols along the tracks correspond to $z=0$, 1, 1.5,  2, 2.8 and 3. The black solid lines show the limits of our modified BM/BX criterion. The distribution of the optically-selected COSMOS galaxies at $1.5<z<2.8$ (black dots, limited for clarity to sources with $i^{+}<25$ and photometric uncertainties smaller than 0.1\,mag in $u^{*}$, $V_{J}$ and $i^{+}$) illustrates the high reliability of our color selection.}
  \label{fig:BM_track_opt}
\end{figure*}

At the wavelengths where these two selection techniques can be applied, the COSMOS observations were carried out at the CFHT and at the Subaru Telescope with the $u^{*}$, $g^{+}$ and $r^{+}$ filters \citep{Capak:07}, which differ substantially from the filters used by \citet{Steidel:04}.  As a result we were not able to adapt the initial $BM/BX$ selection to the COSMOS photometry by applying simple terms of color corrections and/or systematic offsets to our data.  Following the approach proposed by \citet{Grazian:07} we considered a slightly modified version of the $BM/BX$ criterion using the $u^{*}-V_{J}$ and $V_{J}-i^{+}$ colors instead of the colors initially used by \citet{Adel:04}. To determine the most appropriate color cuts for our $u^{*}V_{J}i^{+}$' diagram, we computed the evolution of color tracks followed by star-forming galaxies as a function of redshift, in a way very similar to that employed by \citet{Adel:04} so as to ensure a selection of star-forming sources as close as possible to the original criterion. We used the spectral energy distributions of two starburst galaxies among the list of galaxy templates considered by \citet{Ilbert:09} for their analysis of the COSMOS photometric redshifts. These two SEDs were chosen as being the most representative templates of galaxies beyond $z~\sim1$ according to the SED fitting infered by \citet{Ilbert:09}. These two templates correspond to starburst models taken from (\citealt{Bruzual:2003}, hereafter BC03) to which emission lines has been added to have a better understanding of the COSMOS colors (for more details see \citealt{Ilbert:09}).
  Following \citet{Adel:04}, we also took into account the effect of extinction by reddening  each template up to $E(B-V)=0.2$ assuming the attenuation curve of \citet{Cal:00}.

The exact determination of  our new selection is illustrated on Fig~\ref{fig:BM_track_opt}. First we  defined the  lower and upper limits of our criterion on the $u^{*}-V_{J}$ color using the location of the galaxy tracks  at respectively $z=1.5$ and $z=2.8$:

\begin{equation}
\begin{split}
&u^{*}-V_{J} > 0.52 (V_{J}-i^{+}) - 0.1\\
&u^{*}-V_{J} < 2.5 (V_{J}-i^{+}) + 1.0
\end{split}
\end{equation}

Second, we used a lower limit of $-0.25$ for the $V_{J}-i^{+}$ color. It is similar to the cut chosen by \citet{Grazian:07}, although we note that its effect is negligible since very few sources are bluer than this limit. 
Finally, we imposed an upper limit on this second color so as to allow the selection of sources corresponding to the reddest templates of our diagram ($E(B-V)$=0.2\,mag):

\begin{equation}
u^{*}-V_{J}  < -12.5 (V_{J}-i^{+}) + 9.0
\end{equation}

Similar to what had been noticed by \citet{Adel:04},  the galaxy tracks illustrated in Fig~\ref{fig:BM_track_opt}
reveal an important degeneracy between the colors of low-redshift galaxies and more distant sources affected by dust extinction. The $V_{J}-i^{+}$ upper limit that we adopted corresponds to the compromise of allowing the selection of dusty galaxies while minimizing the contamination of sources at $z<1$. As we will see later in this Section, this will have strong impact on the selection of luminous and dusty star-forming objects with $E(B-V)>0.2$ using the $BM/BX$ technique. We kept however a strict cut at $E(B-V)=0.2$ not to alter the original $BM/BX$ criterion proposed by \citet{Adel:04}. 

To verify the robustness of our new color selection we overploted in Fig~\ref{fig:BM_track_opt}
the observed colors of the $i^+$--band selected COSMOS galaxies with the aforementioned redshift cut of $z>1.5$
\citep{Capak:07,Ilbert:09}. For clarity, only sources with $i^{+}<25$ are shown, so as to minimize the dispersion coming from photometric uncertainties. As expected from the SED tracks, we see that 
our criterion is globally well suited to the identification of optically-selected star-forming galaxies at $1.5<z<2.8$. Out of the 28\,141 $i^+$--band  sources located in this redshift range and selected at  $i^{+}<25$  from the photometric redshift catalog of \citet{Ilbert:09}, 24\,826 sources (i.e., 88\%) satisfy our modified BM/BX selection.  Among the remaining objects that escape the selection, some are substantially affected by dust extinction and their optical colors are thus redder than allowed by the criterion, while some others are characterized by  photometric uncertainties moving them just outside of our BM/BX selection box. Most of them, however, correspond to blue sources with photometric redshifts just above $z=1.5$. Therefore the small incompleteness of our criterion mostly reflects  the color degeneracy that appears at this redshift because of SED variations and dust extinction, as well as uncertainties affecting our photometric redshifts in the COSMOS   field. Relaxing our lower limit on the $u^{*}-V_{J}$ to reduce this incompleteness of 12\% would inevitably increase the fraction of contaminants from sources at $z<1.5$.

\subsubsection{Distant  ``stellar bump dominated'' galaxies}

After the successful launch of the {\it Spitzer Space Telescope}, the advent of deep surveys carried out with the IRAC instrument enabled a new approach for identifying distant star-forming galaxies, based on the rest-frame 1.6$\mu$m bump produced by the H$^-$ opacity minimum in the atmospheres of cool stars \citep{Simpson:99,Sawicki:02}. At $1.5<z<3$ the signature of this stellar bump is shifted between 3 and 9$\mu$m and it reveals itself through specific colors across the IRAC bands. It allows the identification of distant galaxies in a way that is less subject to extinction compared to the optically-based selections discussed earlier.

Several selection techniques have been explored to apply this method to large galaxy samples. They involve either a direct fit of the IRAC fluxes with stellar templates  \citep[e.g.,][]{Sorba:10}, the use of two-color criteria \citep[e.g.,][]{Papo:08,Huang:04,Huang:09} or the simple identification of the SED peak across the 4~IRAC channels  \citep[e.g.,][]{Lonsdale:09}. For our current analysis we considered the selection introduced by \citet{Huang:04}.  It is based on the color cuts $0.05 < [3.6] - [4.5] < 0.4$ and $-0.7< [3.6] - [8.0]< 0.5$, where [3.6] denotes the AB magnitude in the 3.6$\mu m$ band (and likewise for the 4.5 and 8.0 $\mu m$ bands).  Extensive follow-up have confirmed the reliability of this criterion for selecting galaxies at $1.5<z<3$ \citep{Huang:09,Desai:09}. These sources will be refered as the IRAC Peakers hereafter.

\subsection{Selection of Optically Faint IR-bright sources at z$\sim$2}

While the selection techniques discussed in the previous section pertain by definition to sources primarily identified at optical and near-IR wavelengths, observations undertaken in the thermal Infrared and the submillimeter have revealed a large number of luminous galaxies associated with optical counterparts much fainter than the typical magnitude limits considered in the surveys of the distant Universe carried out in the visible \citep[$I \sim 25$\,mag AB, e.g.,][]{Hughes:98,Houck:05,Yan:05}.  Given the extreme IR-to-optical colors that result from these faint magnitudes at visible wavelengths, various criteria based on the mid-IR/optical flux ratio have thus been proposed for selecting dusty star-forming sources at $z>$\,1. However the relevance of the galaxy sub-samples selected with this approach and their contribution with respect to the global population of high redshift sources has barely been explored so far. Here we  describe the selection technique that was proposed by \citet{Dey:08}, based on the ratio between the 24$\mu$m and the optical $R$-band flux densities. Using the $Spitzer$ observations of the Bootes Field and the NOAO Deep Wide-Field Survey (NDWFS) they analyzed the population of 24$\mu$m sources brighter than 0.3\,mJy (i.e., the depth of the MIPS imaging in Bootes) and they defined a selection of dust-enshrouded high redshift sources\footnote{refered as ``Dust-Obscured Galaxies'' in their selection at $F_{24\mu m} > 0.3$\,mJy.} with the following criterion:

\begin{equation}
R - [24]  \geqslant 7.4~mag~(AB)
\label{eq:DOG}
\end{equation} 

where [24] refers to the AB magnitude measured in the MIPS-24$\mu$m band\footnote{The relation given by \citet{Dey:08} uses the Vega system and it was converted following the MIPS Data Handbook: $[24](Vega) = -2.5 \times log_{10} \frac{f_{24\mu m} (Jy)}{7.14}$.}.  Using spectroscopic follow-up performed with the IRS spectrograph on-board $Spitzer$ and with the LRIS/DEIMOS instruments at the $Keck$ telescopes, they have shown that their technique provides a reliable selection of dusty luminous galaxies at $1.5<z<3$, with a small fraction of contaminants from lower redshifts.

\section{Application to the 24$\mu$m selected galaxy populations in COSMOS}

\subsection{The optical/near-IR selections applied to the 24$\mu m$ galaxy population: associated redshift distributions}

To quantify how the UV/optical selections of distant sources suffer from biases and incompleteness due to dust extinction, we applied the various criteria previously discussed to the MIPS-24$\mu$m sample described in Sect.\,\ref{sec:dat}. We then analyzed the relative weight of the different sub-samples of 24$\mu$m sources identified with the $BzK$, BM/BX and IRAC peaker selections with respect to the whole population of 24$\mu$m-selected galaxies.
 
\begin{figure*}[htbp] 
   \includegraphics[width=0.5\textwidth]{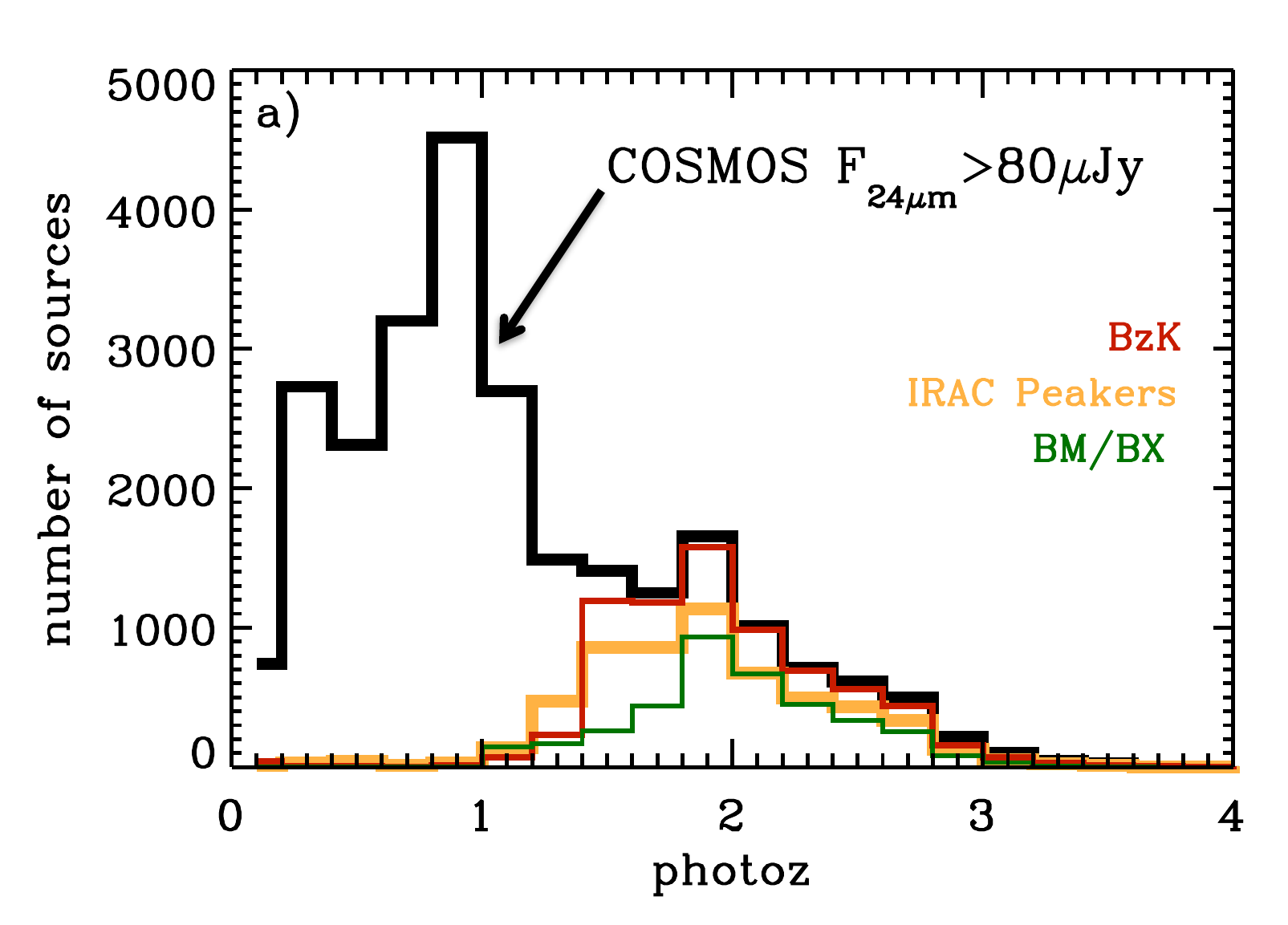}
     \includegraphics[width=0.46\textwidth]{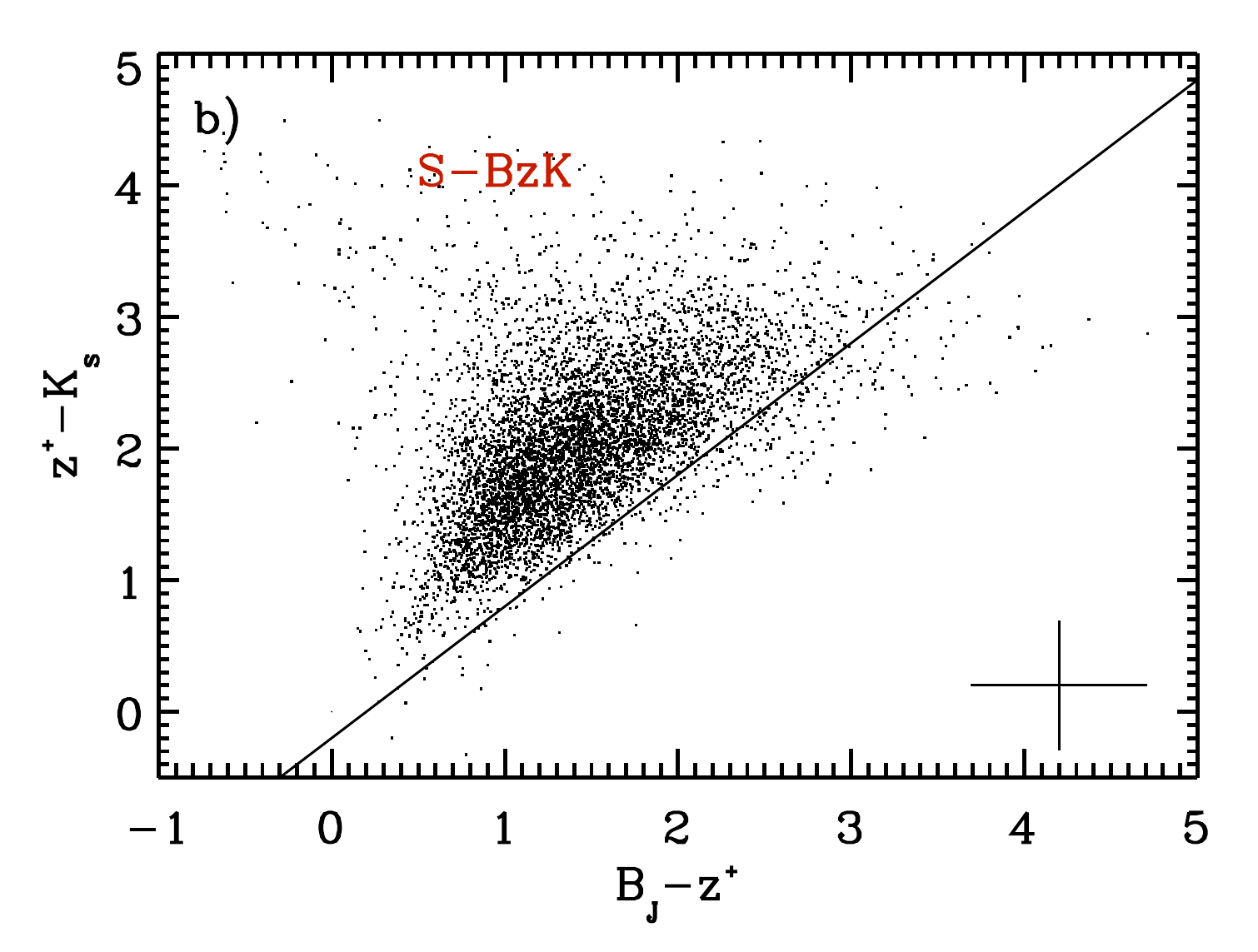}
    \includegraphics[width=0.5\textwidth]{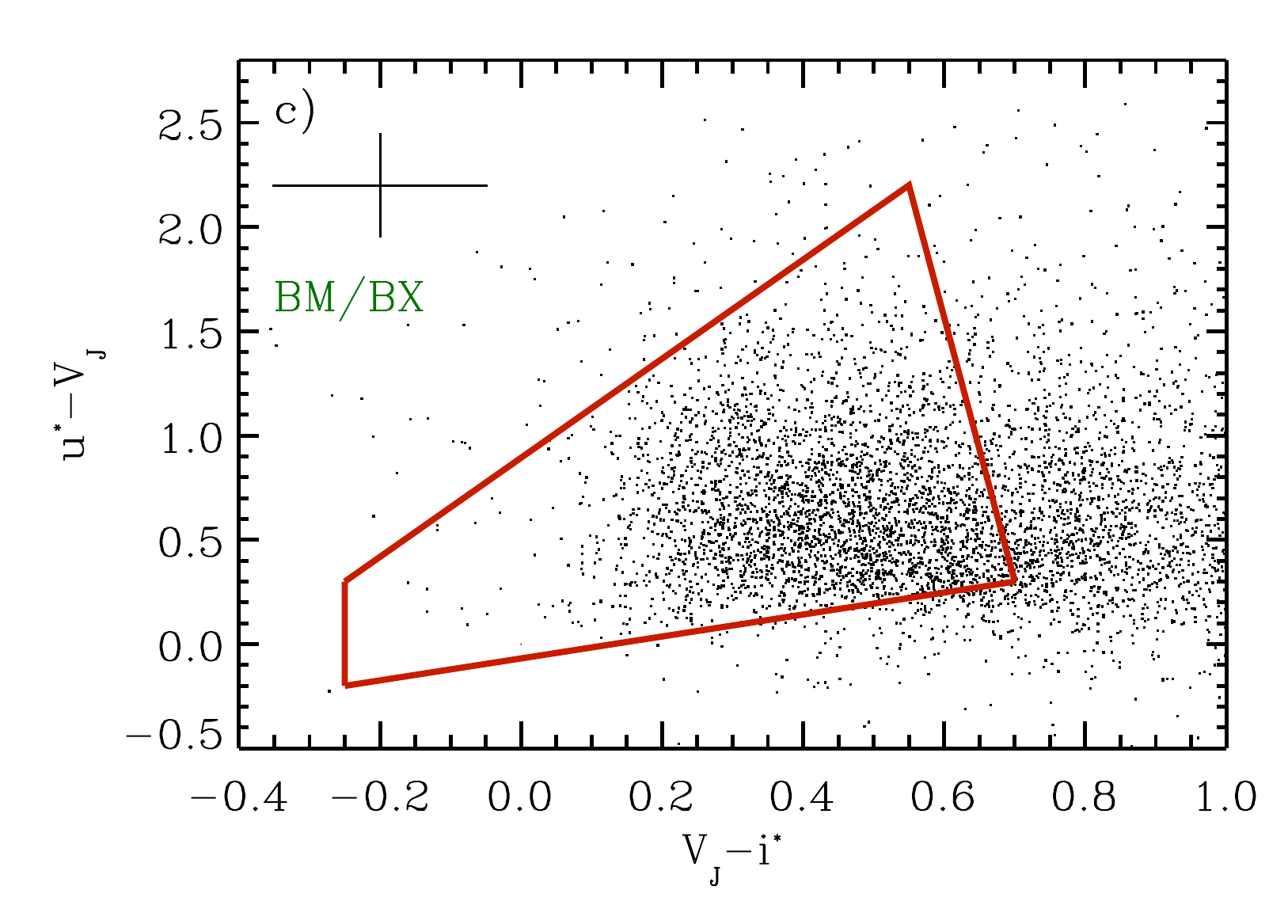}
    \includegraphics[width=0.5\textwidth]{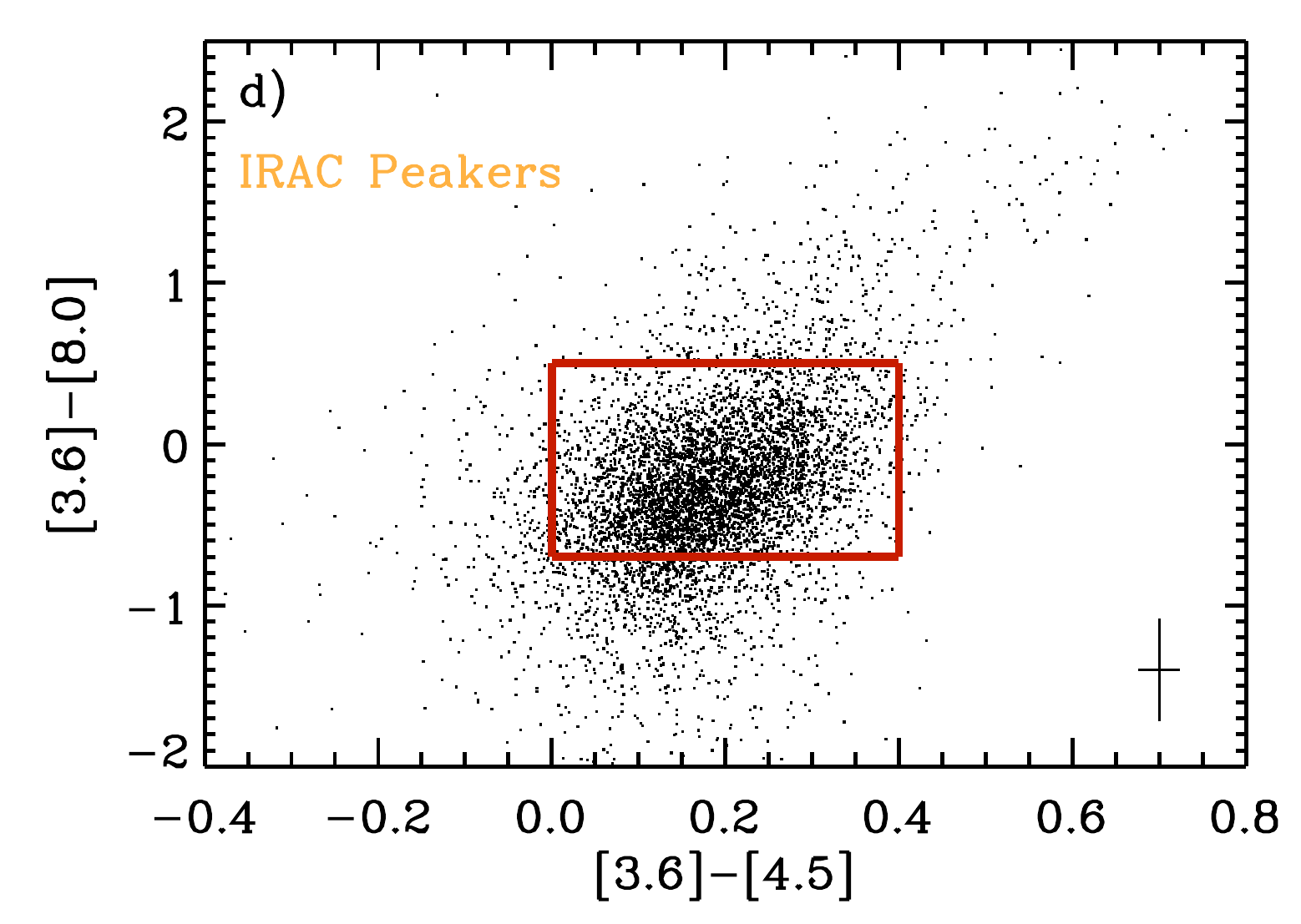}
    \caption{(a) Redshift distributions of the MIPS galaxies identified with the $BzK$ (red), BM/BX (green) and IRAC peaker (orange) selection techniques, compared to the distribution obtained for the whole sample of COSMOS sources with $F_{24\mu m} > 0.08$\,mJy (black solid histogram).  (b)~The $BzK$ color distribution of the MIPS galaxies identified at $1.4<z<2.5$ with the COSMOS photometric redshifts. The $BzK$ star-forming sources (S-BzK) are identified above the diagonal solid line.  (c)~Distribution of the $1.5<z<2.8$ MIPS galaxies  in the $V_J-i^+$ versus $u^* - V_J$ color diagram. The area  defined by the solid line corresponds to our modified BM/BX criterion (see Sect.\,3.1.2). (d)~IRAC colors of the MIPS sources located at $1.5<z<3$. The box illustrates the IRAC peaker selection criterion proposed by \citet{Huang:04}. The typical uncertainties for each population are shown in the corners of the figures.
}
    \label{fig:selec}
\end{figure*}

Our results are illustrated in the different panels of Fig.\,2.  On one hand, Fig.\,\ref{fig:selec}a shows the redshift distributions associated with our three UV/optical selections and compared to the redshift distribution obtained for the entire 24$\mu$m source population in COSMOS. On the other hand, Figs.\,\ref{fig:selec}b, \ref{fig:selec}c \& \ref{fig:selec}d represent the corresponding color diagrams along with the distribution of the 24$\mu$m sources independently identified in the redshift range associated to each selection ($1.4<z<2.5$ for the $BzK$ selection, $1.4<z<2.8$ for our modified BM/BX criterion, and $1.5<z<3$ for the IRAC peakers) using our redshift identifications from the catalog of \citet{Ilbert:09}. The uncertainties characterizing the photometric redshifts of the 24$\mu$m galaxy population in COSMOS have been thoroughly discussed by \citet{Emeric:09} and \citet{Ilbert:09}. Even at $1<z<3$ they are substantially smaller than the typical redshift range where the color selections operate (1$\sigma_z < 0.15$ up to $i^+_{AB} < 25$\,mag and $z<3$), and therefore they should not affect our global conclusions.  Not surprisingly we see that the distributions characterizing the $BzK$, BM/BX and IRAC peaker sub-samples globally cover the redshift range of $1<z<3$, as expected from the definition of their associated selections. Nonetheless, these distributions clearly show that some criteria present substantial biases. Hereafter we discuss in more details the redshift properties characterizing each individual sub-sample.

\begin{itemize}
\renewcommand{\labelitemi}{$\bullet$}
\item{\underline{\it {BzK galaxies:}}}   
Out of the 7\,227 MIPS sources lying in the redshift range $1.4<z<2.5$, we found that 6\,623 objects satisfy the $BzK$ criterion. This represents a fraction of 92\% of the whole MIPS population at these redshifts, implying that the $BzK$ selection is particularly efficient and weakly affected by the effect of dust obscuration. This high efficiency can be explained by the fact that in the $BzK$ diagram the extinction vector evolves parallel to the line defining the $BzK$ criterion for star-forming galaxies ($Bzk=-0.1$), hence allowing the selection of deeply obscured sources provided they are detected in the $K$--band. Furthermore, the redshift distribution (Fig.\,2a) reveals that  the presence of contaminants at $z<1.4$ is clearly negligible, while the $BzK$ colors of the MIPS sources identified in the $1.4<z<2.5$ redshift range  (based on photometric redshifts) but  not selected as $BzK$ sources are distributed very close below the threshold of $BzK = -0.1$ (see Fig.\,2b). It shows that the small incompleteness of the $BzK$ selection at $1.4<z<2.5$ could simply result from  optical/near-IR photometric uncertainties spreading sources below the borderline of the criterion, as well as small  uncertainties on the photometric redshifts of sources close to the redshift boundaries of the selection. 
  As an interesting aside we also note that 84\% of the MIPS sources with $2.5<z<3$ are still selected as $BzK$ sources. It suggests that the $BzK$ criterion could also  be applied successfully at even higher redshift, at least for our sample of dusty luminous galaxies.
\\
 It is important to stress that the high success rate of identifications that we find with the $BzK$ technique is largely due to the depth of the optical and near-IR observations available in the  COSMOS field. A large number of high redshift 24$\mu$m galaxies are indeed associated with faint optical/near-IR counterparts and the ratio of MIPS sources identified with the $BzK$ selection would have been much lower if we had used shallower data at optical and near-IR wavelengths. As an example we restricted the COSMOS $K_s$--band sample to respectively $K_s$\,=22, $K_s$\,=22.5 and $K_s$\,=23, and we found that  with these additional cuts only 42\%, 72\% and 89\% of the MIPS population at $1.4 < z < 2.5$ would have been identified with the $BzK$ selection. Given the large contribution of dusty luminous galaxies to the cosmic star formation density at $z \sim 2$  \citep[e.g.,][]{Caputi:07,Rodi:10} this clearly illustrates the critical need for deep near-IR observations when probing the star-forming high-redshift galaxy population with the $BzK$ criterion.
\\
Finally, we also recall that 4\% of sources from the very first sample of 24$\mu$m detections extracted from the MIPS imaging of COSMOS (see Sect. \ref{sec:dat}) could not be matched with the optical catalog of \citet{Ilbert:09}. We then have no information on their colors and their redshift. Given their faintness at short wavelengths these galaxies most probably lie at z$>$1. If they were all located at $1.4<z<2.5$ they would represent 15\% of the whole population of 24$\mu$m sources identified in this redshift range. The fraction of MIPS sources selected as BzK galaxies would thus decrease from 92\% down to $\sim$78\%.
\\
\item{\underline{\it {BM/BX sources:}}} The distribution shown in Fig~\ref{fig:selec}a reveals that the BM/BX criterion enables the identification of dusty galaxies in the redshift range expected from the analysis that we presented in Sect.\,3.2, but with a much lower efficiency than the $BzK$ and the IRAC peaker selections. Out the 7\,459 MIPS sources with $1.5< z < 2.8$, only 3\,754 (i.e., $\sim$50\%) are indeed identified as BM/BX galaxies. To understand the origin of this bias we illustrate in Fig. \ref{fig:selec}c the distribution of the 24$\mu$m sources at $1.5< z < 2.8$ in the BM/BX diagram. It clearly shows that a large fraction of them is located outside of the BM/BX selection area that we defined earlier, because of much redder colors than observed on average in the optical sample selected at $1.5< z < 2.8$. Over this redshift range, the comparison between the BM/BX properties of the MIPS sources and that of the COSMOS $i^+$-band selected population displayed in Fig.\,\ref{fig:BM_track_opt} is in fact particularly striking.  According to the galaxy tracks represented in this figure, the MIPS sources located to the right hand side of the BM/BX selection box correspond either to galaxies with substantial extinction ($E(B-V) > 0.2$) or to lower redshift contaminants that we could have falsely identified with high-redshift objects.  The later is however unlikely given the very low fraction of catastrophic failures among the photometric redshifts of the MIPS-selected population in COSMOS \citep{Ilbert:09,Emeric:09}, as well as the excellent agreement that we just found between the $BzK$ selection and the MIPS sample at $z \sim 2$. Therefore we conclude that the large fraction of 24$\mu$m sources missed by the BM/BX selection correspond to galaxies truly located at $1.5< z < 2.8$ but strongly reddenned by dust extinction. This is obviously not surprising since the $Spitzer$ mid-IR observations naturally favor the selection of dusty objects.  Yet it clearly reveals how the BM/BX criterion is biased toward the identification of UV-bright sources with no or little dust obscuration ($E(B-V) < 0.2$)
 and how this selection can thus  miss a large amount of luminous dust-enshrouded galaxies in the distant Universe. In Sect.5 we will quantify in more details the impact of this bias on the bright end of the IR luminosity function of star-forming galaxies at $z \sim 2$. 
\\
\item{\underline{\it {IRAC peakers:}}}
  Among the 7\,755 MIPS sources located at $1.5<z<3$ a total of 4\,942 galaxies were selected as IRAC peakers. This corresponds to an average completeness of $\sim$64\% over this redshift range, showing that this criterion is also suited to distant galaxies but slightly less efficient than the $BzK$ selection. This smaller efficiency must be due to other effects than dust extinction since the IRAC criterion is based on data taken at longer wavelengths. A possible explanation is that the color selection of IRAC peakers has been restricted to identify only the galaxies where the stellar bump is particularly pronounced, while dusty objects usually show a wider diversity of SEDs in the near-IR given the combined effect of stellar and hot dust emission.  Besides, the COSMOS photometric uncertainties in the IRAC bands are typically larger than those obtained at shorter wavelengths \citep[e.g.,][]{Ilbert:09}. As we can see in Fig.\,2d this may result in distributing galaxies at $1.5<z<3$ over a slightly wider range of IRAC colors while increasing the contamination from galaxies at lower redshifts. Indeed we note that the redshift distribution of the 24$\mu$m sources identified as IRAC peakers extends below $z \sim 1.5$ with 23\% of the sub-sample being located at $1<z<1.5$.  Finally, although bright X-ray AGNs were removed from our initial sample, we can not fully exclude a residual contamination from high-redshift obscured nuclei detected at 24$\mu$m. Their SED is usually characterized by a rising feature-less hot dust continuum, which can totally outshine the stellar emission of the galaxy when the AGN is luminous enough \citep{Brand:06,Menendez:09,Desai:09}.  In this case their host exhibit much redder colors than those allowed by the ``IRAC peaker'' selection, which would contribute to lowering its efficiency.
  However, the AGN contribution to the whole population of IR galaxies and to the Cosmic Infrared Background is likely not larger than 15\% \citep[e.g.,][]{Jauzac:11}. Hence, their possible contamination in our sample can not be the main explanation for the incompleteness of the IRAC peaker method.
\end{itemize}

\subsection{Infrared-luminous sources with optically-faint counterparts at z$\sim$2 }

\begin{figure}[htbp] 
    \includegraphics[width=0.49\textwidth]{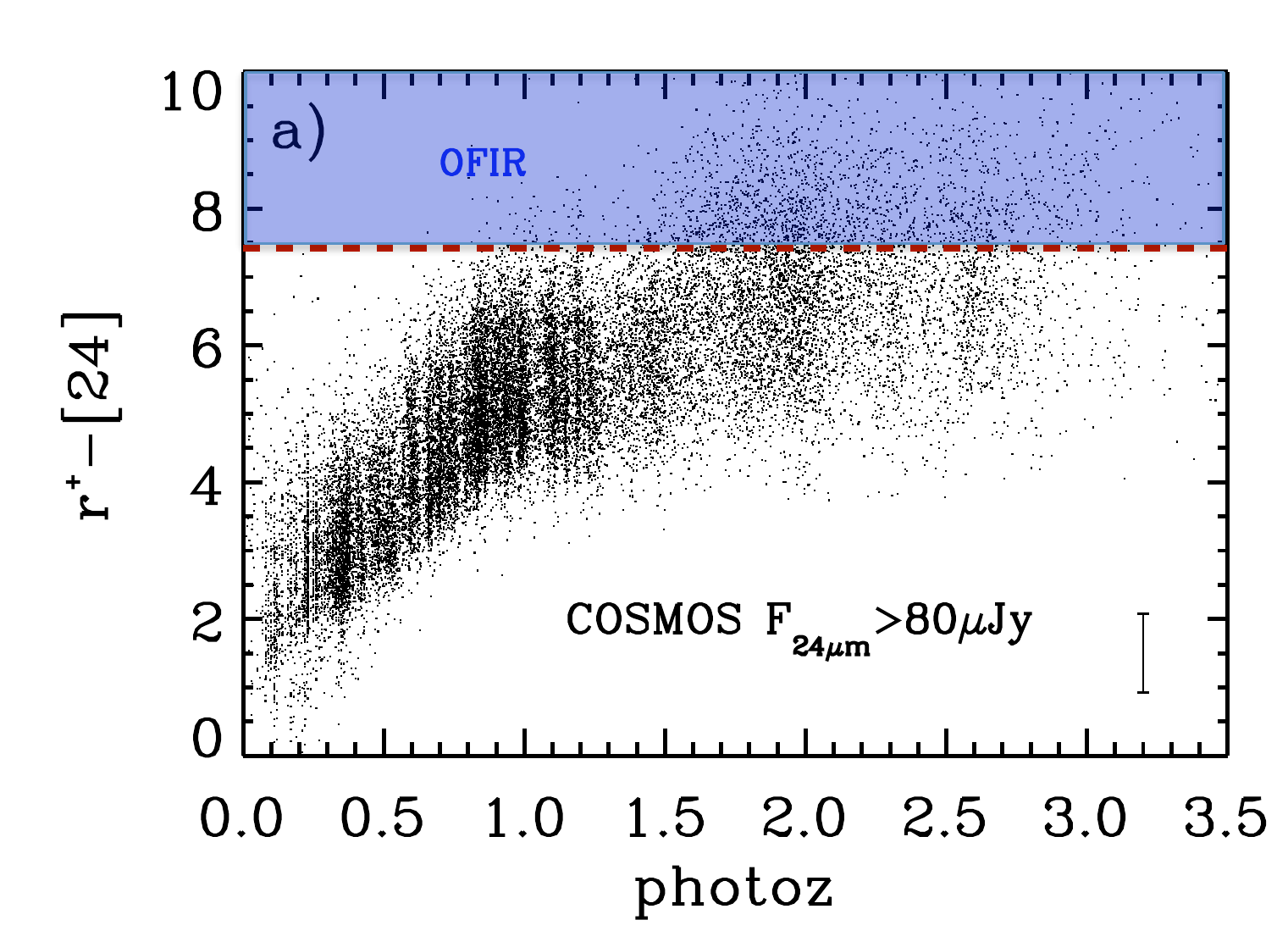}
    \includegraphics[width=0.49\textwidth]{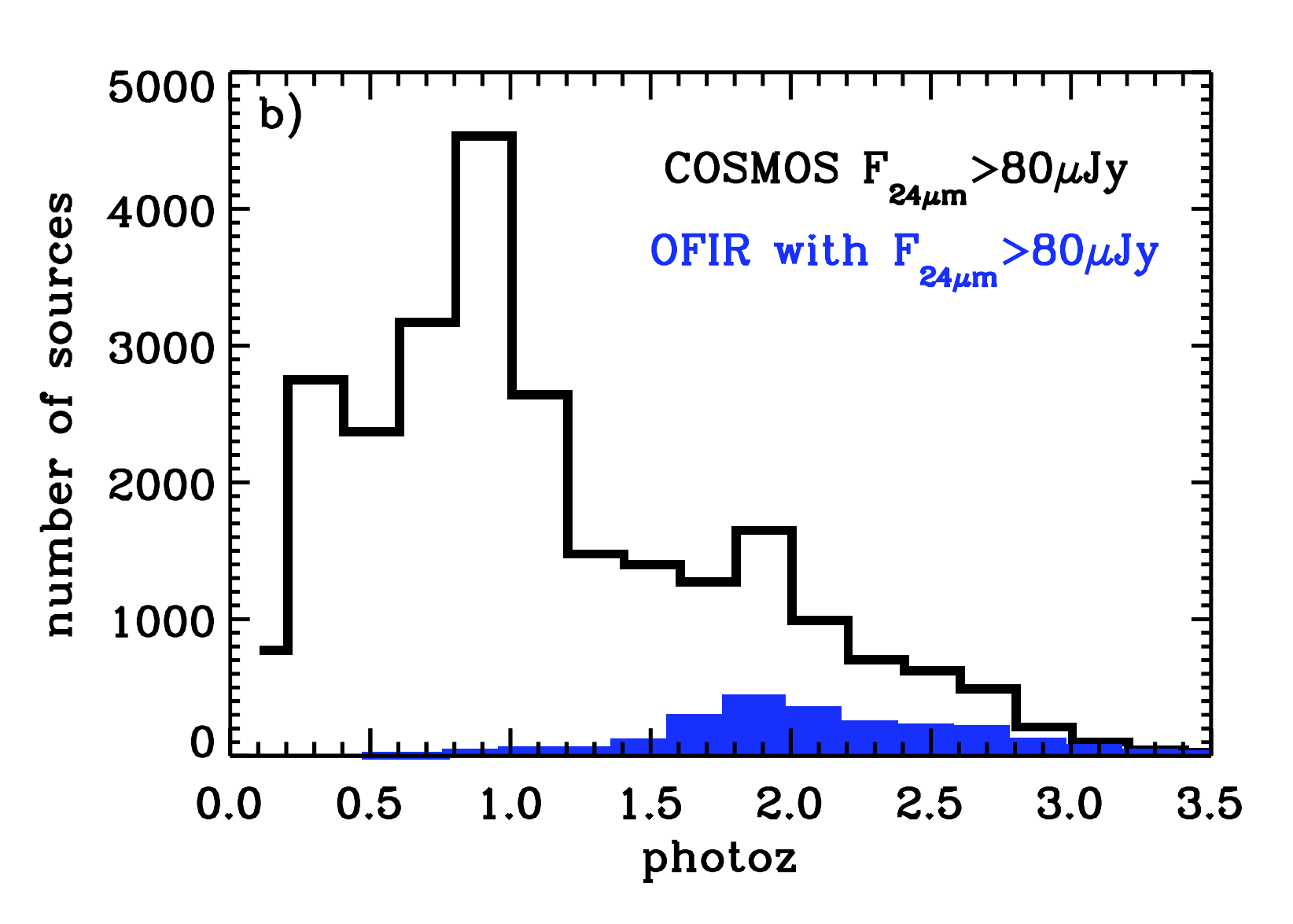}
    \includegraphics[width=0.49\textwidth]{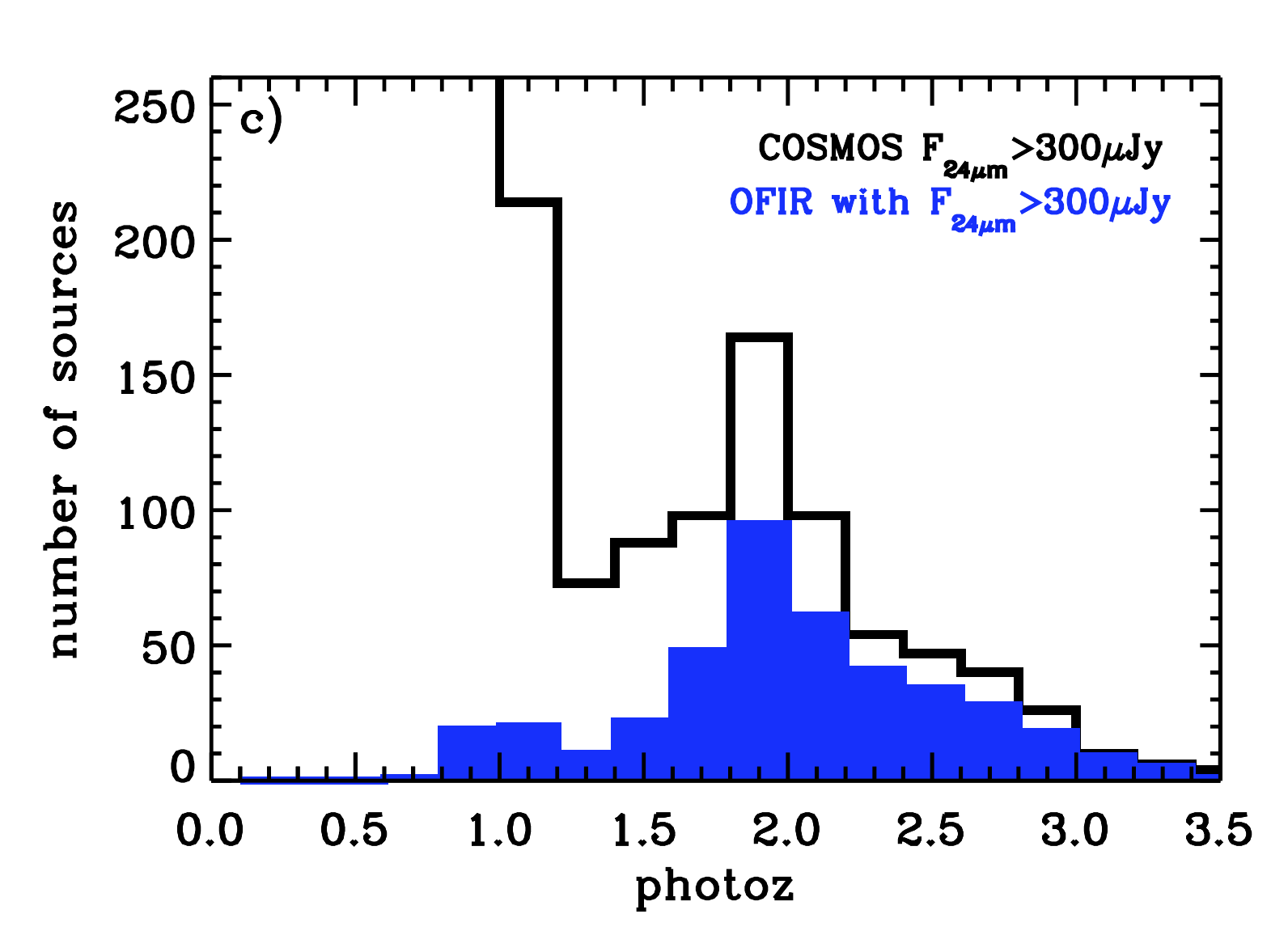}
   \caption{(a) The $r^+-[24]$ color distribution of the MIPS 24$\mu$m sources as a function of redshift in the COSMOS field.  Galaxies selected as Optically-Faint IR-bright sources are characterized by $r^+-[24]$ colors redder than the threshold indicated by the dashed line. This criterion effectively selects galaxies at $1.5<z<3$. The typical uncertainties on the $r^+-[24]$ color have been calculated at $1.5<z<3$ and are shown on the right corner. (b)~Redshift distribution of the Optically-Faint IR-bright galaxies with $F_{24\mu m} > 0.08$\,mJy (blue filled histogram) compared to the redshift distribution obtained for  the full MIPS-selected galaxy population in COSMOS (black solid line). (c)~Redshift distributions derived from the MIPS sample restricted to  sources with  $F_{24\mu m} > 0.3$\,mJy (same colors definitions than (b)). The relative contribution of the OFIR galaxies at $1.5<z<3$ increases with mid-IR luminosities.
}
    \label{fig:histo_DOG1}
\end{figure}

In the analysis presented in the previous sub-section we have seen that a substantial fraction of the high-redshift mid-IR selected galaxies are strongly reddened by dust extinction, and their identification can thus be missed when purely relying on rest-frame UV color selection techniques. Conversely, the selection of distant sources characterized by a large excess of IR emission such as the Optically-Faint IR-bright (OFIR) objects will necessarily be biased {\em against\,} dust-free galaxies. To quantify this effect, we applied the optical/mid-IR criterion described in Sect.\,3.2  to our MIPS galaxy sample using the $r^+$-band observations of COSMOS and we extended this criterion down to $F_{24\mu m}$\,=\,0.08\,mJy in order to probe a larger range of mid-IR luminosities. In Fig.\,\ref{fig:histo_DOG1}a we show the distribution of the $r^+ - [24]$ colors for the MIPS-selected sources as a function of redshift, which confirms that most of the Optically-Faint IR-bright galaxies satisfying the criterion of \citet{Dey:08} are indeed located at $1.5<z<3$.  Figs.\,\ref{fig:histo_DOG1}b \&\,\ref{fig:histo_DOG1}c illustrate the corresponding redshift distributions obtained for MIPS sources with respectively $F_{24\mu m}>0.08$\,mJy and $F_{24\mu m}>0.3$\,mJy, compared to the distribution of the full MIPS-selected sample in COSMOS.  The Optically-Faint IR-bright sources only represent $\sim$30\% (2\,350 sources on the 7\,755 sources with $1.5<z<3$) of the population of MIPS galaxies selected with $F_{24\mu m} > 0.08$\,mJy at $1.5<z<3$, but as we will see in the next section their contribution is rising with the mid-IR luminosity. Above a 24$\mu$m flux of 0.3\,mJy their fraction reaches $\sim$60\% of the 24$\mu$m-selected population at $z \sim 2$.

The fact that the fraction of galaxies with 24$\mu$m to optical-band flux ratios satisfying Eq.~\ref{eq:DOG} rises as a function of 24$\mu$m flux is not an artificial selection effect due to the relative depths of our MIPS and $r^+$-band COSMOS observations. Even at the 24$\mu$m flux limit of 0.08\,mJy the optical COSMOS data are deep enough to identify galaxies as red as the color threshold used in the criterion proposed by \citet{Dey:08}. Also it implies that the 4\% of the initial 24$\mu$m detections that we could not identify at optical and near-IR wavelengths (Sect.2) satisfy Eq.~\ref{eq:DOG} independently of their mid-IR flux density. Since these extremely red colors are unlikely associated to galaxies at $z<1$ (see Fig.\,\ref{fig:histo_DOG1}a) the relative contribution of galaxies selected at $1.5<z<3$ with this criterion could reach $\sim$45\% at $F_{24\mu m} > 0.08$\,mJy.

\begin{figure}[htbp] 
   \includegraphics[width=0.49\textwidth]{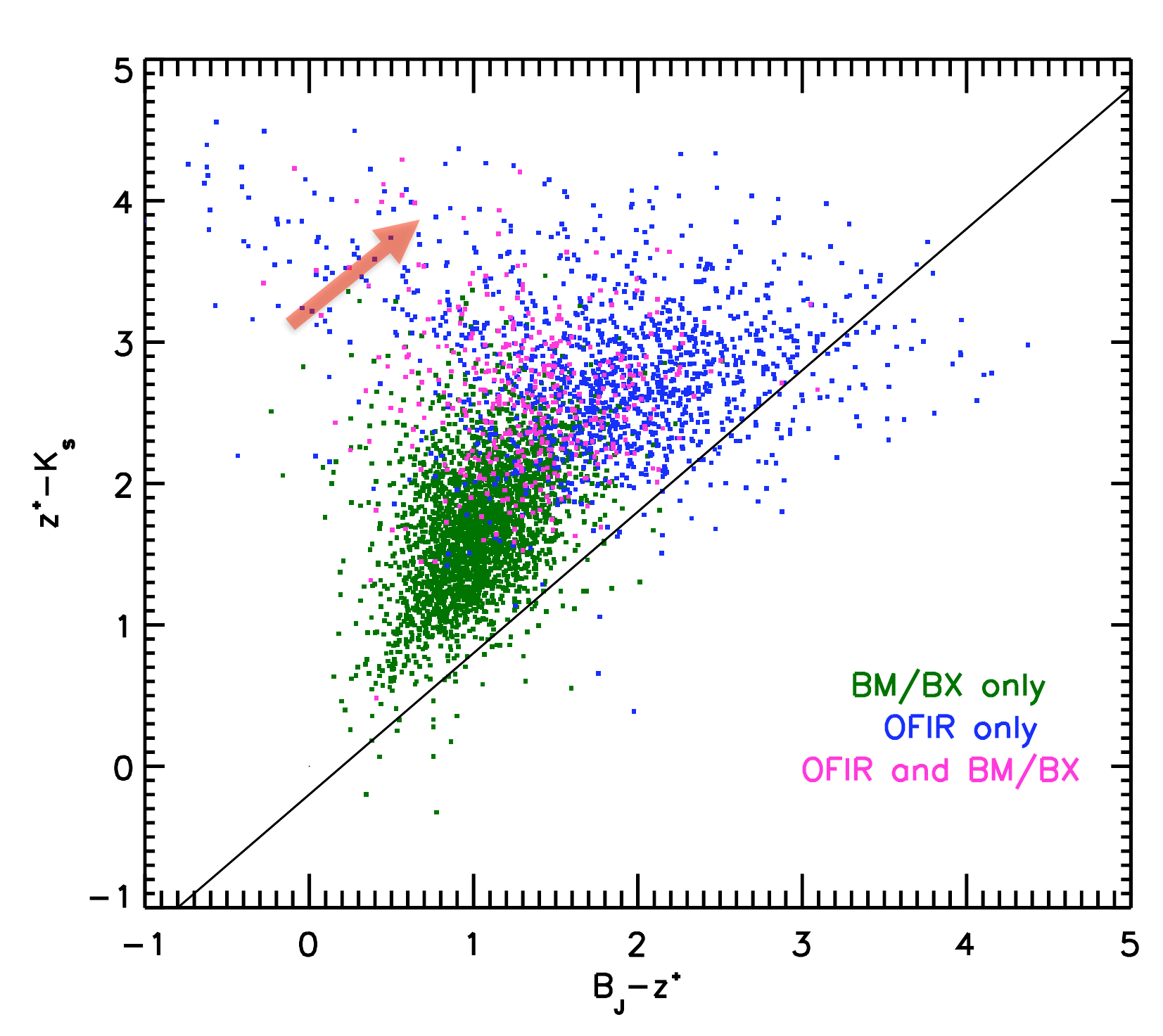} 
   \caption{Distribution of the 24$\mu$m sources selected as otically-faint IR-bright galaxies (blue dots) and with the BM/BX criterion (green dots) in the $BzK$ diagram. Only sources at $1.5<z<2.8$ are represented and galaxies satisfying the two selections are shown with pink color. The Optically-Faint IR-bright sources are characterized by redder colors, most likely due to higher dust extinction. The red arrow represents the extinction vector.}
   \label{fig:BzK_diagram}
\end{figure}

As we already argued this selection of high-redshift galaxies based on extremely red mid-IR/optical colors is by construction biased toward dust-obscured sources with very faint optical luminosities. For example we found that 66\% of these  objects have $r^+>26$\,mag while the $r^+$--band magnitude distribution of the MIPS sources with $1.5 < z < 2.8$ peaks at $r^+ \sim 25$\,mag (only 21\% of them have $r^+>26$\,mag). Although they do not represent a dominant population in terms of source density, this nicely illustrates the contribution of luminous high-redshift star-forming galaxies that may be missed by optical surveys because of dust obscuration. In fact, 65\% of the Optically-Faint IR-bright sources have an extinction of $E(B-V) \ge 0.4$ according to the optical SED fitting of \citet{Ilbert:09}, while this percentage decreases to only 44\% for the whole population of MIPS sources in the same redshift range. Furthermore, galaxies selected with this technique differ dramatically from those identified with the BM/BX selection, which we found to be strongly biased $against$ dusty high-redshift galaxies. In Fig.\,\ref{fig:BzK_diagram} we  compare in the $BzK$ diagram the distribution of the Optically-Faint IR-bright sources with the distribution of galaxies selected with our modified BM/BX criterion. The overlap between the two sub-samples is relatively small as only 8\% of the MIPS sources at  $1.5 < z < 2.8$ satisfy the two selection criteria, and not surprisingly the Optically-Faint IR-bright objects are characterized by much redder colors than the BM/BX sources. Since the Optically Faint IR-bright galaxies and the BM/BX selected sources share very similar redshift distributions (compare Figs.\,2a \&~3b) and given the  extinction vector at $z \sim 2$ in the $BzK$ diagram (Fig.\,\ref{fig:BzK_diagram}), the lack of overlap between these two populations is mostly due to the effect of dust extinction.

\begin{figure*}
  \centering
   \includegraphics[width=0.95\textwidth]{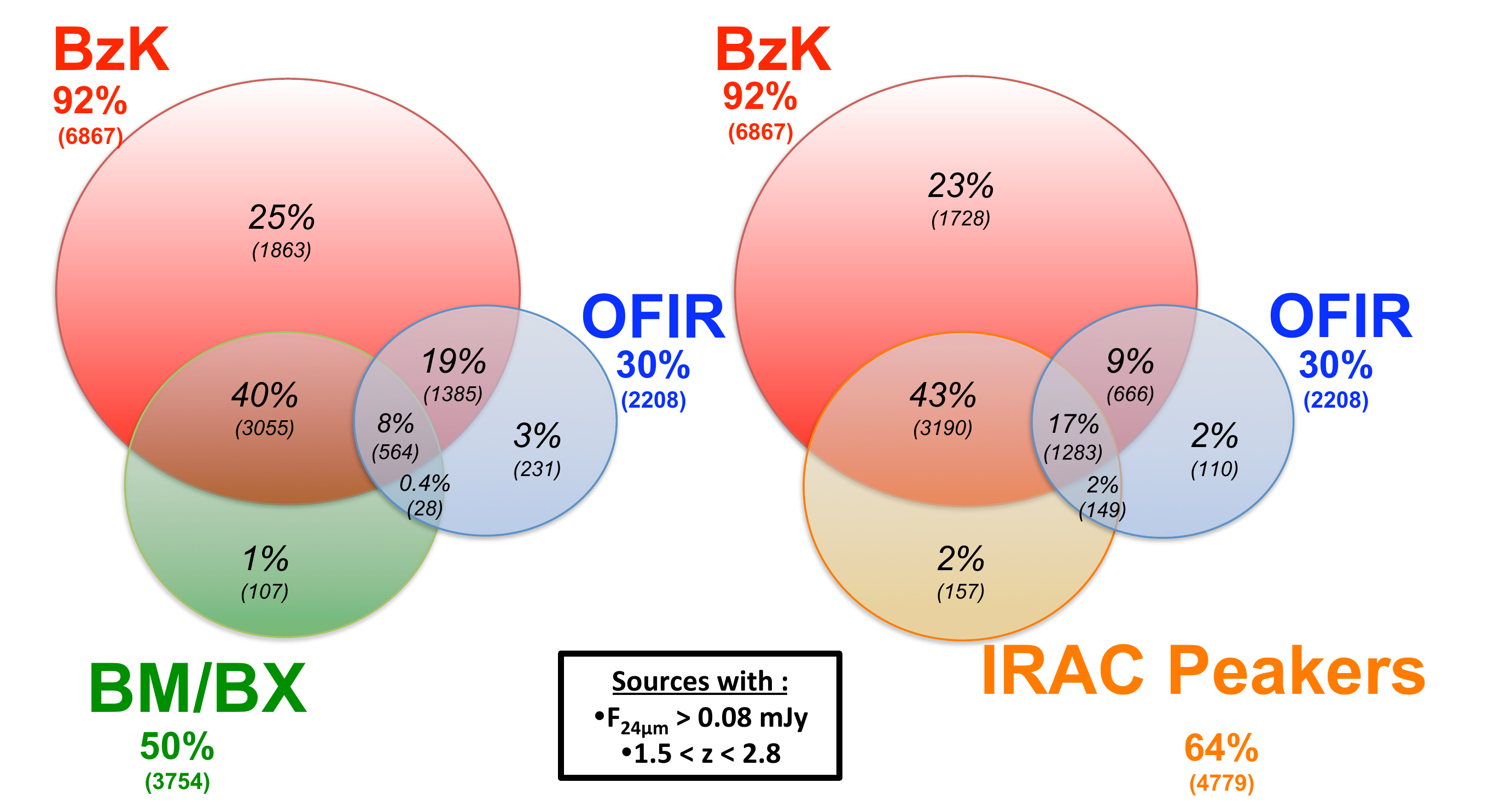}
   \caption{Overlap between the populations of 24 $\mu m$ sources with  $F_{24\mu m} > 0.08$\,mJy on the redshift range $1.5<z<2.8$. The percentages in bold are related to the total number of sources selected by the criterion considered, the percentages in italic correspond to the overlap between the populations.
   {\it Left:\,} Overlap between the BzK, the BM/BX and the Optically Faint IR-bright sources. {\it Right:\,} Overlap between the BzK, the Optically Faint IR-bright sources and the IRAC Peakers.} 
   \label{fig:overlap}
\end{figure*}

In Fig.\,\ref{fig:overlap} we illustrate in a more quantitative way how the Optically Faint IR-bright objects and the other selections overlap with each other. Since we found that the high efficiency of the $BzK$ criterion is maintained up to $z \sim 3$ (at least when applied to our mid-IR sample, see Sect.\,4.1), we compared our different selections over the largest possible redshift range  (i.e., $1.5 < z < 2.8$) to minimize the statistical uncertainties. More than half of the population of Optically Faint IR-bright sources are also selected as IRAC Peakers indicating that an important part of the sources presenting extreme MIR color excess have an SED dominated by the stellar bump.

\section{Galaxy 8$\mu$m luminosity function and Star Formation Rate Density at $z \sim 2 $}
\label{sec:LF}

In the two previous sections we described a number of color selection techniques that have been widely used for identifying star-forming galaxies at $1.5<z<3$, and we quantified their respective contribution to the total number density of MIPS-selected high-redshift sources in order to estimate the bias that these selections suffer because of dust extinction.  We want to extend this analysis by measuring the mid-IR luminosity function of the different sub-samples of galaxies selected based on these techniques, so as to infer their contribution to the total IR luminosity density of the Universe and the cosmic star formation rate density observed at $z \sim 2$. Here we detail the methods used to obtain the 8$\mu$m rest-frame Luminosity Function (LF), the IR luminosity density and the Star Formation Rate (SFR).

\subsection{8$\mu$m luminosity function}

The galaxy luminosity functions were all computed at rest-frame 8$\mu$m. The redshift distributions of the sources selected with the techniques described earlier mostly peak at $z \sim 2$ (Figs.\,2a \&~3b) and at these redshifts the measure of 8$\mu$m luminosities ($L_{8\mu m}$) based on 24$\mu$m fluxes is therefore almost independent of the SED templates assumed for deriving the $k$-corrections.  Also, the 8$\mu $m luminosity of star-forming galaxies mostly originates from the emission of the 7.7$\mu$m and 8.6$\mu$m Polycyclic Aromatic Hydrocarbon (PAH) lines, which represent the most prominent features in the mid-IR spectrum of dusty galaxies \citep[e.g.,][]{Laurent:00,Dale:01,Brandl:06,Smith:07}. These features are stochastically heated by the radiation field of young stellar populations, and as it was demonstrated by numerous studies of local sources, their luminosity is tightly correlated with the total Infrared luminosity of galaxies, and therefore with their SFR \citep[e.g.,][]{Roussel:01,Wu:05,Brandl:06,Cal:07,Diaz:08,Goto:11}. Furthermore, observations of distant sources with {\it Spitzer\,} and more recently with {\it Herschel} have revealed that this correlation between the 8$\mu $m and the total IR luminosity of star-forming galaxies extends up to at least $z \sim 2$ (\citealt{Pope:08,Rigby:08,Bavouzet:08,Menendez:09}, Elbaz et al., submitted), suggesting that the 8$\mu$m emission is a fairly good tracer of star-forming activity also in high-redshift galaxies.

To convert the observed MIPS-24$\mu$m fluxes into luminosities at 8$\mu$m we used the library of IR SED templates of \citet{Chary:2001}. In this library the galaxy IR SEDs vary as a function of total IR luminosity ($L_{\rm IR} = L_{8-1000\mu m}$)
and for a given redshift the flux density measured at 24$\mu$m  corresponds to a unique monochromatic luminosity at any IR wavelengths. However we stress again that for our current analysis the $k$-corrections only depend on the shape of these templates between $\lambda_0 = 8\mu$m and $\lambda_1 = 24\mu$m/(1+$z$). At $z \sim 2$  they are thus barely sensitive to the choice of SEDs and do not depend on the uncertainties  that have been shown to affect the extrapolations of mid-IR fluxes to total IR luminosities \citep{Elbaz:10}. In fact we computed the rest-frame 8$\mu$m luminosities of the MIPS galaxy sample using the libraries of IR SED templates proposed by \citet{Dale:2002}, \citet{Lagache:04} and \citet{Rieke:2009}, and we found virtually no difference with the results obtained with the library of \citet{Chary:2001}. This effect will be discussed in more details in a forthcoming paper by Le Floc'h et al. (in prep.) on the evolution of the mid-IR luminosity function in COSMOS.

\begin{figure}[htbp] 
  \includegraphics[width=0.5\textwidth]{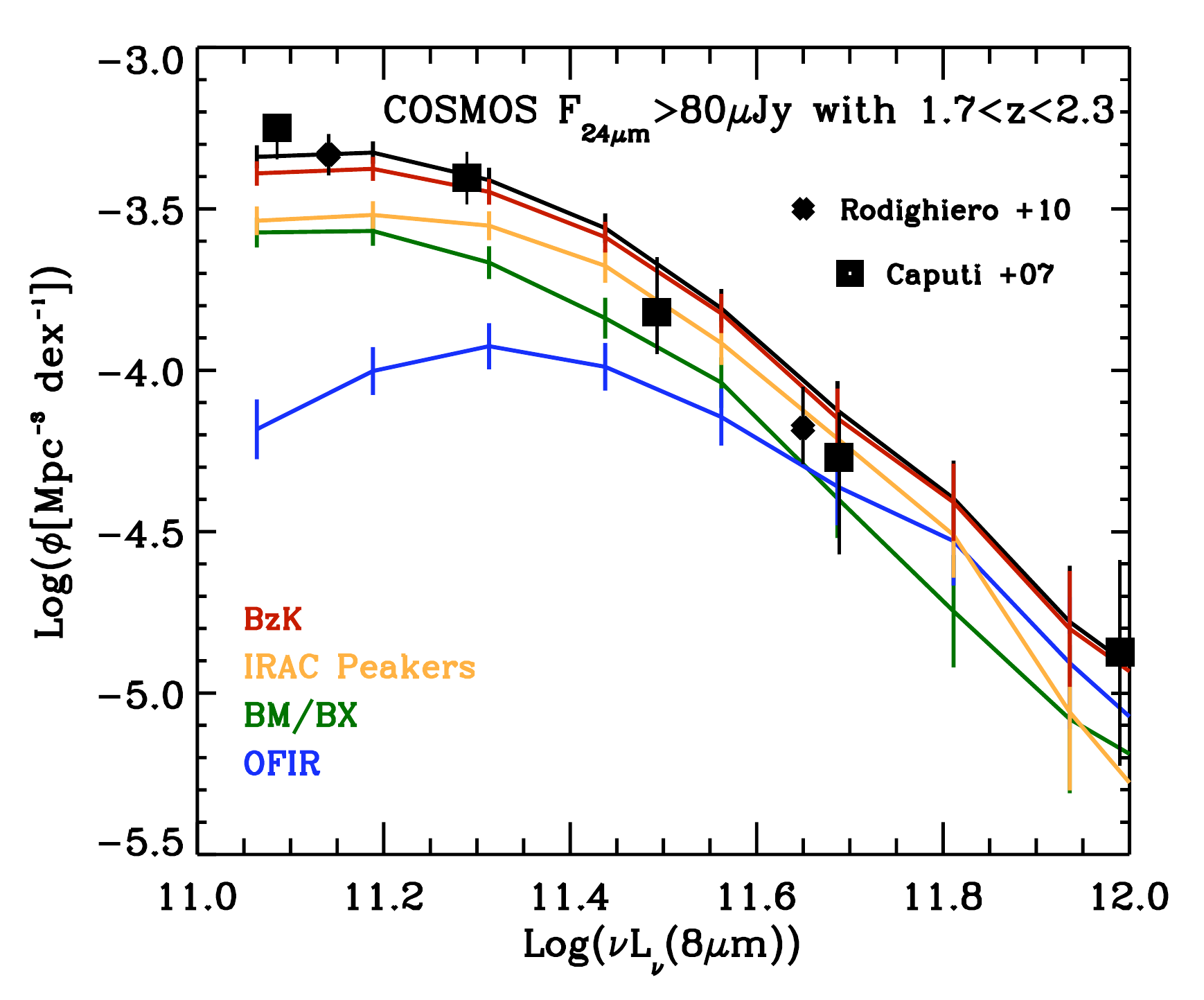}
   \includegraphics[width=0.49\textwidth]{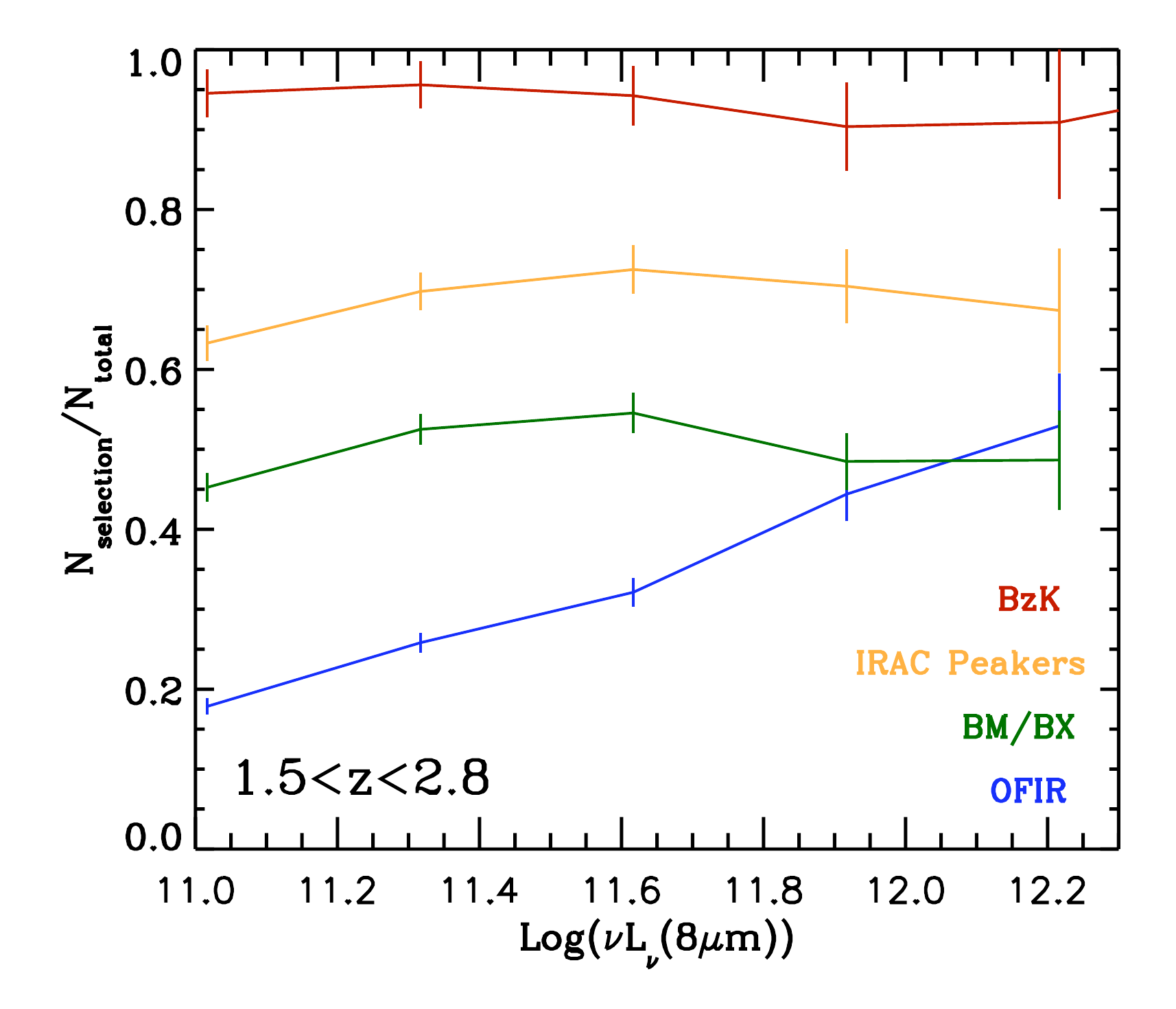}
   \caption{{\it Top:\,} The 8$\mu$m luminosity functions of MIPS galaxies identified at $1.7<z<2.3$ with the color selection techniques described in Sects.\,3 \&~4 (red: $BzK$; green: BM/BX; yellow: IRAC peakers; blue: Optically-Faint IR-bright galaxies) compared to the total 8$\mu$m luminosity function measured in the COSMOS field (black solid line).  Previous measurements of the total 8$\mu$m LF at $z\sim2$ by \citet[filled diamonds]{Rodi:10} and \citet[filled squares]{Caputi:07}  are also shown for comparison. Error bars were estimated from the Poisson noise. {\it Bottom:\,} Fraction of galaxies identified with each of the color selections as a function of $L_{8\mu m}$ and in the $1.5<z<2.8$ redshift range. Color coding is similar to that used in the top panel.} 
   \label{fig:LF}
\end{figure}

Based on these 8$\mu$m luminosities we computed  the luminosity functions (LF) associated with our different sub-samples over the $1.7<z<2.3$ redshift range  and we compared these results to the global LF obtained at the same redshifts from the full sample of MIPS sources in COSMOS.  These luminosity functions are illustrated in the top panel of Fig~\ref{fig:LF}. They  were estimated using the 1/V$_{max}$ method \citep{Schmidt:68}, which   is advantageous in two ways: no hypothesis on the shape of the LF is needed and the luminosity function is directly measured from the observations.  For each object we estimated the maximum comoving volume where it can be detected within our given redshift bin ($1.7<z<2.3$) using $V_{max} = V_{zmax} - V_{zmin}$, where z$_{min}=1.7$   and z$_{max}$ is the smallest value between   the redshift bin upper limit ($z=2.3$) and the redshift until which the source could be observed considering the 24$\mu$m flux limit of our survey (i.e., $F_{24\mu m} = 0.08$\,mJy). The uncertainties on the determination of the LFs were estimated from the Poisson noise associated to the number of sources detected in each  bin of luminosity. Since we are only interested in quantifying the relative contribution of  galaxies identified with the color selection techniques discussed earlier as a function of mid-IR luminosity, we neglected the effect of other uncertainties like the 24$\mu$m flux errors and the cosmic variance.

\subsection{Star Formation Rate Estimates and SFR density}

Because the mid-IR emission of star-forming galaxies originates from dust features heated by ionizing photons coming from young stars, their mid-IR luminosity correlates pretty well with their star-formation rate  \citep[e.g.,][]{Roussel:01,Forster:04,Wu:05}. For example \citet{Cal:07} and \citet{Diaz:08} analyzed the mid-IR properties of HII regions taken from a sample of local star-forming sources. They established a significant correlation with a logarithmic slope $\sim$\,1 between their 8$\mu$m luminosity surface density corrected for stellar contribution and their luminosity measured from the Pa$\alpha$ emission line corrected for extinction. Similarly, the combination of the mid-IR and far-IR observations of the nearby Universe conducted with IRAS, ISO, $Spitzer$\, and AKARI revealed a tight relationship between $L_{8\mu m}$ and the total IR luminosity of local galaxies \citep[e.g.,][]{Bavouzet:08,Goto:11}.  The deepest observations of the sky undertaken with $Herschel$\, have recently shown that this correlation extends up to at least $z \sim 3$ (Elbaz et al. submitted) and it may thus represent a universal property shared among star-forming galaxies at all cosmic times.

The relation obtained by Elbaz et al. (2011, submitted) between $L_{8\mu m}$ and the total Infrared luminosity is linear ($L_{\rm IR}/L_{8\mu m} \sim 4.9 \pm 0.2$\,dex). Assuming the standard conversion inferred  by \citet{Kennicutt:98}  between the  $L_{\rm IR}$  and the  star formation rate  ($SFR ~ [M_{\odot} yr^{-1}] = 1.72 \times 10^{-10} L_{IR}[L_{\odot}]$), we then obtain:

\begin{equation}
SFR  (M_{\odot}\,yr^{-1}) = 8.4^{+4.9}_{-3.1} \times 10^{-10} L_{8\mu m}(L_{\odot}) 
\label{eq:sfr1}
\end{equation} 

\noindent At the typical 8$\mu$m luminosities found in our high-redshift sample (i.e., $L_{8\mu m} \sim 10^{11}-10^{12}L_{\odot}$), this is fully consistent with the relation we would get by combining the correlation measured by \citet{Goto:11} with  the calibration from \citet{Kennicutt:98}:

\begin{equation}
SFR  (M_{\odot}\,yr^{-1}) = (34\pm9) \times 10^{-10} \times (L_{8\mu m} / L_{\odot})^{0.94}
\label{eq:sfr2}
\end{equation} 

\noindent Similarly, it agrees with the conversion we would obtain by applying the relationship from \citet{Cal:07} to a typical star-forming disk (i.e., 2--5\,kpc in diameter) and assuming a case B recombination for converting the Pa$\alpha$ luminosity into a star formation rate ($SFR ~ [M_{\odot} yr^{-1}] = 6.79 \times 10^{-41} L_{\rm Pa\alpha}[erg~ s^{-1}]$, \citealt{Oster:89,Alonso:06}). This last approach relies  on the assumption made for the size of the emitting region, but in  the range of 8$\mu$m luminosities that we measured this effect is overwhelmed by the internal dispersion found in our sample and the uncertainty affecting the slope of the correlation between the Pa$\alpha$  and the 8$\mu$m luminosity surface densities (\citealt{Cal:07,Diaz:08}, see also \citealt{Alonso:06}). This relation  is also  consistent with the Pa$\alpha$  and the 8$\mu$m luminosities that were recently measured in a submillimeter lensed galaxy at $z \sim 2.5$ by \citet{Papo:09}. It suggests that it could still be applied to high-redshift sources even though it was derived from  observations of the nearby Universe.

\begin{figure*}[htbp] 
   \centering
 \includegraphics[width=0.7\textwidth]{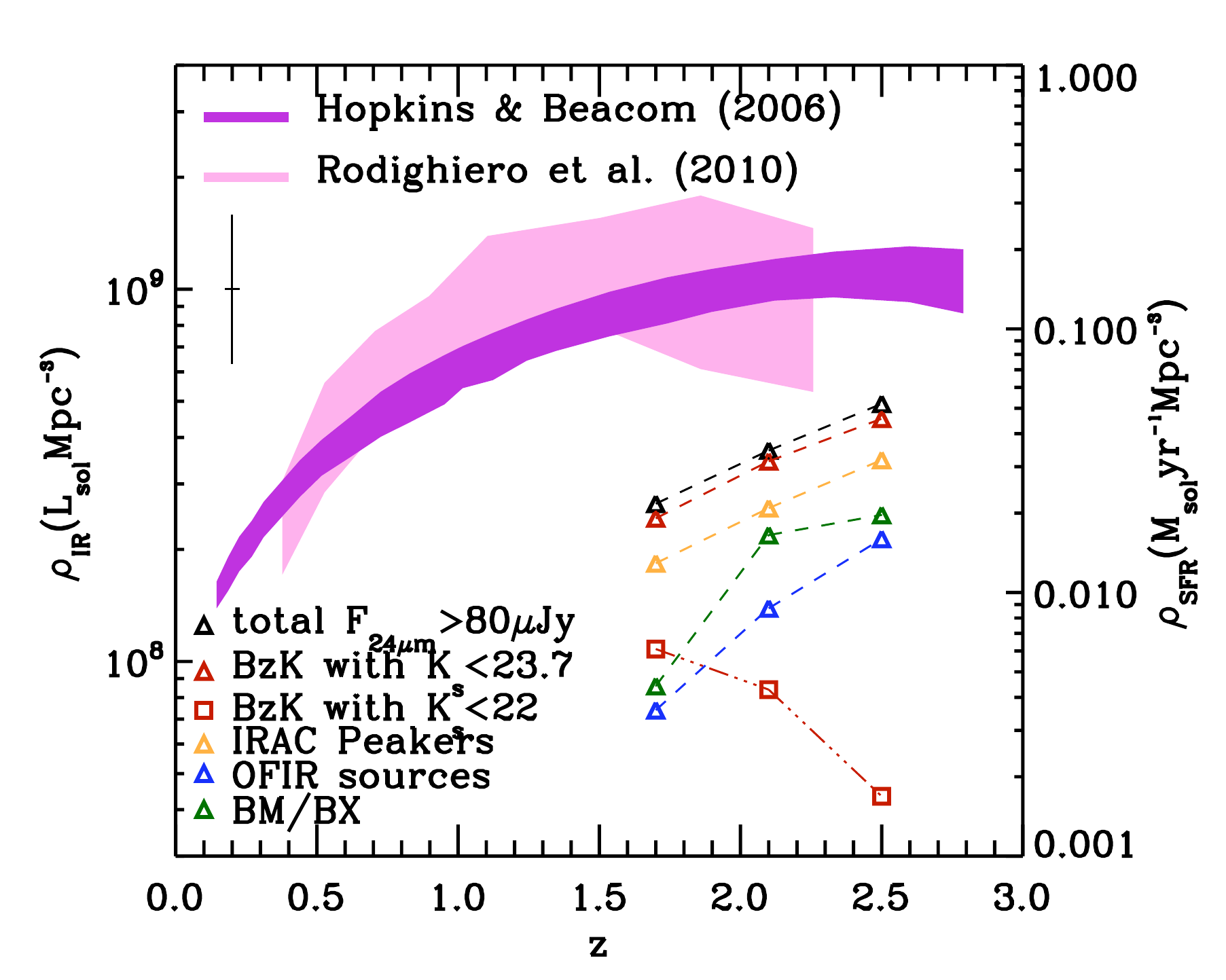} 
   \caption{Total Infrared Luminosity Density produced at respectively $1.5<z<1.9$, $1.9<z<2.3$ and $2.3<z<2.7$ by 24$\mu$m sources with $F_{24\mu m} > 0.08$\,mJy (black triangles), compared to the contribution of 24$\mu$m sources identified with the selection techniques described in Sects.3 \&~4 (red: $BzK$; green: BM/BX; yellow: IRAC peakers; blue: Optically-Faint IR-bright galaxies with photo-z identification). For galaxy samples where the MIPS 24$\mu$m emission is mostly produced by star formation (i.e., $BzK$ and BM/BX soursces, IRAC peakers) the IR Luminosity Density can be read as a Star Formation Density (right vertical axis) assuming the conversion from \citet{Kennicutt:98}. For comparison, the evolution of the total IR luminosity density inferred by \citet{Rodi:10} up to $z \sim 2.5$ is illustrated by the light shaded region. The cosmic star formation history derived by \citet{Hopkins:06} is represented by the dark shaded area. The typical uncertainty affecting our estimates is shown by the vertical line in the top left corner of the diagram.}
      \label{fig:rho_SFR}
\end{figure*}

Given the good agreement between  these different relationships, we   thus inferred the total IR luminosity and  star formation rate of the COSMOS MIPS sources using the linear relation obtained by Elbaz et al. (2011, submitted) as well as our Eq.\,\ref{eq:sfr1}, baring in mind that the other possible methods would have led to very similar results. 
Based on these determinations of $L_{\rm IR}$ and SFR, we computed the contributions of the different populations of MIPS-selected sources discussed in Sect.\,3 to the IR luminosity density of the Universe and the cosmic star formation rate density  in three redshift bins: $1.5<z<1.9$, $1.9<z<2.3$ and $2.3<z<2.7$.  To reach this goal we derived luminosity functions in a way similar to the method described in Sect.\,5.1, and we integrated the LFs above the luminosity corresponding to the 24 $\mu$m flux limit of our survey ($F_{24\mu m} = 0.08$\,mJy) at the median of  each redshift bin. Since we only aim at determining how the various galaxy sub-samples $detected$\, with MIPS contributed to the total SFR density, we did not assume any extrapolation of the MIPS-selected source population to the faint end of the luminosity function. Our results are shown in Fig.\,\ref{fig:rho_SFR}, which also illustrates  the evolution of the total IR luminosity density inferred by \citet{Rodi:10} up to $z \sim 2.5$ and  the cosmic star formation history compiled by \citet{Hopkins:06}. The 1306 sources with no optical counterparts are obviously not taken into account here but we already saw that they contribute only a small fraction of the total number of galaxies identified in the COSMOS field. Furthermore, most of these unidentified sources have low 24$\mu$m fluxes close to the sensitivity limit of the COSMOS MIPS data. This implies that their contribution to the IR luminosity density is even less than their contribution to the number of mid-IR selected sources, and therefore the lack of identification for these sources can not severely bias our results.

\section{Discussion}
\label{sec:discussion}

\subsection{Contributions to the IR luminosity function}

The top panel of Fig.\,\ref{fig:LF} illustrates the luminosity functions associated with the MIPS sources identified with the four color selection techniques described in Sect.\,3, compared to the total $8\mu$m luminosity function observed at $1.7<z<2.3$ in the COSMOS field. The later is also compared with the $8\mu$m LFs obtained by \citet{Caputi:07} and \citet{Rodi:10} from the GOODS and SWIRE surveys, which show relatively good agreement with our estimates. 
Except for the selection of Optically-Faint IR-bright galaxies, our analysis seems to indicate that the fraction of MIPS high-redshift galaxies selected with the $BzK$, BM/BX and IRAC peaker criteria does not critically depend on the $8\mu$m luminosity itself. To better illustrate this result we reproduced the evolution of these fractions with $L_{8\mu m}$ in the bottom panel of the figure. To increase the statistics we extended the redshift bin up to $1.5<z<2.8$, which corresponds to the largest redshift range where the efficiencies of our different color selections can be simultaneously compared with one another. We see again that the fractions of MIPS galaxies selected as $BzK$ and BM/BX sources or as IRAC peakers are fairly constant over the range of 8$\mu$m luminosities probed by the MIPS observations. They correspond to the average fractions we had already derived based on their number density and  their redshift distribution. 

While the absence of correlation between $L_{8\mu m}$ and the fraction of $BzK$ sources or IRAC peakers is not necessarily unexpected, the result that we find for the BM/BX galaxies may deserve further explanations. Indeed we attributed the bias observed in the BM/BX selection to the effect of dust extinction (see Fig.\,2c). Given the global trend that exists between galaxy bolometric luminosities and the IR/UV luminosity ratio \citep[e.g.,][]{Bell:03}, one could have expected some broad correlation between the 8$\mu$m luminosity and the fraction of MIPS high-redshift sources missed by the BM/BX criterion. In the most luminous galaxies though, the reddening of the optical light does not correlate anymore with the dust obscuration measured by the excess of L$_{\rm IR}$ over L$_{\rm UV}$ \citep[e.g.,][]{Goldader:02,Reddy:06}.  This can be understood, for instance,  if on one hand the UV/optical emission originates from spatially-extended regions of star formation where dust grains follow a rather clumpy distribution, while on the other hand the IR light predominantly comes from very compact and optically-thick regions close the center of galaxies.
Since the MIPS observations of COSMOS are only sensitive to the very bright end of the high-redshift  galaxy luminosity function, we are likely in this luminosity regime where the $E(B-V)$ extinction is not directly correlated with the level of  star-forming activity, hence explaining the lack of trend between the IR luminosity and the detectability of galaxies based on their $U_nGR$ colors.  Finally, it is also possible that our result at the highest luminosities (L$_{8\mu m} > 10^{12} L_\odot$) is slightly affected by some AGN contribution, which can become significant at bright 24$\mu$m fluxes \citep[e.g.,][]{Martinez:05,Houck:05,Brand:06}. In particular, the blue continuum arising from unobscured type-1 quasars could potentially bias the BM/BX selection. However the majority of the most IR-luminous AGNs in COSMOS have  also been detected in the X-rays \citep{Hasinger:07,Brusa:10}. They have been removed from our initial sample and therefore they should not influence the various trends that we find in  Fig.\,\ref{fig:LF}.

Contrary to the behavior characterizing  the three rest-frame UV/optical selections, the fraction of Optically-Faint IR-bright galaxies is clearly rising with IR luminosity. This is consistent with what we had qualitatively inferred from the comparison of their redshift distributions at $F_{24\mu m}>0.08$\,mJy and $F_{24\mu m}>0.3$\,mJy (see Fig.\,3). In the previous section we argued that these sources are biased toward highly-obscured galaxies. Assuming that dust extinction must correlate with the observed  24$\mu$m/optical flux ratio of galaxies at $z \sim 2$, it could thus explain the trend that we find with luminosity. Furthermore, this trend  could partly originate from an increasing contribution of obscured AGNs to the mid-IR luminosities of galaxies as a function of 24$\mu$m flux \citep[e.g.,][]{Brand:06,Menendez:09}. This rising contribution would boost their mid-IR emission with respect to their optical luminosity, hence leading to an $r^+-[24]$ color satisfying the selection criterion of this population.

\subsection{Implication on the cosmic star formation history}

Even though we did not account for the contribution of galaxies below our 24$\mu$m flux limit we see on Fig.\,7 that 
star-forming galaxies detected in the COSMOS mid-IR survey contribute  a substantial fraction of the total IR luminosity density at $z \sim 2$, in agreement with previous findings \citep[e.g.,][]{Caputi:07,Rodi:10}.  It implies that the incompleteness which affects  the different color selection techniques previously described can result in non-negligible biases with respect to the global picture of galaxy evolution inferred from complete samples of high-redshift sources. For example we find that the MIPS sources selected with the BM/BX criterion contribute  only $\sim$50\% to the IR luminosity density produced by the galaxies with $F_{24\mu m} > 0.08$\,mJy at $z \sim 2$. Similar to what is observed for the other color selections, this fraction is roughly constant over the whole $1.5<z<2.8$ redshift range.  In the case of the BM/BX technique we attributed this bias to the effect of dust extinction,  and since the obscuration globally decreases with bolometric luminosity \citep{Reddy:06} the global incompleteness affecting this selection could probably be smaller if we had considered a deeper 24$\mu$m selection. However, it is interesting to note that a very similar result was also independently obtained  by \citet{Guo:09} based on a semi-analytical approach and using the updated versions of the {\it Millenium Simulations} \citep{Springel:05}. According to the models they found that a large fraction of the total star formation density at $z \sim 2$ originates from galaxies that are too heavily obscured by dust to be selected with the BM/BX criterion. In principle, incompleteness corrections can  be applied to account for the difference between the properties of galaxies satisfying this color selection  and the properties  characterizing the global underlying population of star-forming sources at these redshifts \citep[e.g.,][]{Reddy:08}. However, our analysis and the results from \citet{Guo:09} reveal that a substantial extrapolation accounting for half of the most actively star-forming sources (i.e., $L_{8\mu m} > 10^{11.1} L_{\odot}$, corresponding to $SFR > 100 M_{\odot} yr^{-1}$) would be required to also take into account the large contribution of dusty luminous galaxies to the cosmic star formation density at $1.5<z<2.5$.

Similarly, we find that if the IRAC peaker selection is probably well suited to the identification of galaxies characterized by a pronounced stellar bump,  it can clearly not provide a complete view on the total contribution of massive star-forming sources to the build-up of stellar mass at high redshift. Over the whole redshift  range where this selection technique can be applied ($1.5<z<3$) up to $\sim$30\% of the total 8$\mu$m luminosity density produced at the bright end of the luminosity function is contributed by massive sources escaping the IRAC color criterion previously discussed.  In the case of the Optically-Faint IR-bright sources we find that this fraction can even go up to  $\sim$70\%.  Although the incompleteness affecting these two selections could be partly explained by the hot dust continuum of  dust-obscured AGNs  contributing  to the rest-frame near-IR emission of the most luminous (i.e., $L_{8\mu m} > 10^{12} L_{\odot}$) galaxies at high redshift \citep[e.g.,][]{Houck:05, Desai:09}, it is unlikely to be the case for the majority of distant sources with more typical luminosities and which are known to exhibit strong PAH features characteristic of star-forming activity at mid-IR wavelengths \citep{Fadda:10}. Furthermore we recall the strong correlation that has been recently established between the 8$\mu$m luminosity and the total IR luminosity of galaxies up to $z \sim 3$ (Elbaz et al., submitted), which reinforces the idea that the mid-IR luminosity characterizing the general galaxy population is mostly powered by star-forming activity.  We thus conclude that  a non negligible fraction of the cosmic star formation density is missed by these two selection techniques.

On the other hand we note that the $BzK$ approach enables an almost complete selection of star-forming galaxies at $z \sim 2$ at least in the range of luminosities that can be probed by our MIPS observations. This high level of completeness is  observed not only for the bulk of the sample but also for the most luminous -- and potentially the most highly obscured -- galaxies. Using the optically-selected catalog of photometric redshifts in COSMOS \citep{Ilbert:09} we verified that the reliability of the $BzK$ color criterion to select star-forming galaxies in the $1.4<z<2.5$ redshift range  remains also valid at fainter luminosities. It  suggests that this technique  provides a quite powerful way to identify galaxies  at this epoch of cosmic history, independently of the possibility to determine accurate photometric redshifts for these sources. 

More globally speaking, the fact that a non negligible fraction of high-z sources can be missed by optical color selections has also been observed by other groups using different kinds of analysis, and this bias likely originates not only from dust extinction but also from the broader diversity of high-redshift galaxy properties than was assumed in the definition of the color criteria that were explored so far. For instance, Ly et al. (private communication \& in prep.) show that surveys of [OII] emitters can pick up a non negligible fraction of $z \sim 1.5$ galaxies that are missed by the $BzK$ and $BM/BX$ techniques.  Similarly, \citet{Caputi:11} found that at $z = 3 - 4$, $\sim$30\% of the massive galaxies (M$\geq$10$^{11}$\,M$_{\odot}$) would be missed by deep optical surveys, underlying the importance of deep near-IR surveys to supplement the identification of high-z galaxies affected by extinction. Based on deep spectroscopy taken as part of the Vimos VLT Deep Survey, \citet{LeFevre:05} finally underlined that magnitude-selected samples provide a much more complete census of the high-redshift galaxy population than color selection techniques.
 All these results converge in the way that the identifications of distant star-forming galaxies based on UV/optical color criteria result in a view on high redshift star formation that can not be fully representative of the global picture of galaxy evolution.  

\section{Summary and conclusions}
\label{sec:ccl}

Using the exquisite multi-wavelength imaging obtained in the  COSMOS field we analyzed the broad-band optical/near-IR colors and the photometric redshift distribution of a complete sample of distant ($1.5<z<3$) luminous galaxies selected at 24$\mu$m with $F_{24\mu m} > 0.08$\,mJy. To quantify how selection effects and incompleteness issues  can bias the different color selection techniques  often used for identifying populations of high-redshift star-forming sources (e.g., $BzK$, BM/BX and ``IRAC peaker" selections, ``Optically-Faint IR-bright" objects), we measured the fraction of 
MIPS-24$\mu$m galaxies satisfying these color criteria within the redshift range where each of them is expected to apply with high efficiency. We also derived reliable estimates of their rest-frame 8$\mu$m luminosities with minimal effects from the $k$-correction, which allowed us to constrain their star-formation rate and their contribution to  cosmic star formation density. Our results can be summarized as follows:

\begin{enumerate}
\item The $BzK$ color criterion  allows an almost complete ($\sim$90\%) identification of  24$\mu$m sources at $1.4<z<2.5$. The rate of these identifications does not change with IR luminosity, which suggests that this technique provides a highly reliable way to identify distant galaxies independently of obscuration. This can be explained by the weak dependence of the $BzK$ variable ($BzK \equiv [z-K] - [B-z]$\,) on the effect of dust extinction at $z \sim 2$. However we emphasize that the success rate of the $BzK$ selection strongly relies on the availability of deep imaging at optical and near-IR wavelengths. With a cut at $K_s = 22$ (AB), the contribution of $BzK$--identified galaxies to the IR luminosity density produced by sources with $F_{24\mu m} > 0.08$\,mJy would fall down to $\sim$40\%.
\\
\item We adapted the original BM/BX criterion to the existing COSMOS optical photometry  using the deep imaging obtained with the $u^*$--band, $V_J$--band and $i^+$--band filters. We found that up to 50\% of the MIPS selected galaxies at $1.5<z<2.8$  are missed by this selection technique. We explain this result by the effect of dust extinction, which more strongly affects the rest-frame UV emission  of galaxies and makes their SED redder than the typical blue colors  characterizing galaxies satisfying the BM/BX criterion. This is consistent with the results  recently obtained by \citet{Guo:09} based on {\it Millenium simulations}  \citep{Springel:05}, who found that a large fraction of the cosmic star formation density probably originates from massive sources that are too heavily obscured by dust to be identified with the UV color selections. At the bright end of the bolometric luminosity function, the extrapolations and incompleteness corrections usually applied to account for this bias  must be therefore quite large and uncertain.
\\
\item Up to $\sim$30\% of the IR luminosity density produced at $z \sim 2$ by sources brighter than 80$\mu$Jy originates from galaxies that fail to be identified thanks to the spectral shape of  their stellar bump feature at rest-frame 1.6$\mu$m (i.e., the so-called ``{\it IRAC peaker\,}" selection). For the brightest galaxies in the  mid-IR, this effect could be explained by an important contribution of dust-obscured AGNs to the near-IR emission observed in the IRAC bands. For the bulk of the population though, it reflects the larger uncertainties affecting the IRAC photometry as well as the wide diversity of near-IR SEDs characterizing  dusty galaxies, due to the combined contribution of hot dust and stellar emission.
\\
\item The selection of Optically-Faint IR-bright sources based on the $R - [24]$ color  is meant to identify dusty galaxies at $1.5<z<3$. While this technique is  efficient at the very bright end of the mid-IR luminosity function where galaxies are characterized by extremely high 24$\mu$m/optical flux ratios, the fraction of sources selected with this criterion  rapidly decreases at fainter luminosities. Assuming the depth of the COSMOS MIPS survey ($F_{24\mu m} > 0.08$\,mJy) we find that only $\sim$25\% of the IR luminosity density of the Universe at $z \sim 2$ is produced by galaxies with such extreme colors.
\end{enumerate}

These results suggest that  color selections of distant star-forming galaxies can be affected by substantial biases and incompleteness, and therefore  they must be used with strong caution. Depending on the spectral features or the SED range that they are supposed to probe, these high-redshift selections are probably less subject to contaminants compared to the use of photometric redshifts with modest quality if the later are strongly affected by catastrophic failures at $z>1$. However we have shown that the identification of distant galaxies solely based on optical/IR color criteria will most often provide an incomplete view on the whole population of high-redshift star-forming sources, hence requiring large and uncertain extrapolations to account for their incompleteness. Complete identifications based on high-quality spectroscopic redshifts or accurate photometric redshifts thus appear the most reliable approach for probing the formation and the evolution of galaxies in a global and cosmological context.
\\
\\
\\
{\it Acknowledgments:}  It is a pleasure to acknowledge the contribution from 
all our colleagues of the COSMOS collaboration. More 
information on the COSMOS survey is available at 
http://www.astro.caltech.edu/cosmos. This work is based on observations made with the {\it Spitzer Space Telescope}, a facility 
operated by NASA/JPL. Financial supports were provided by 
NASA through contracts nos. 1289085, 1310136, 1282612, and 
1298231 issued by the Jet Propulsion Laboratory. We also want to warmly thank our referee for his/her critical review of the manuscript, as well as 
 Tanio D{\'i}az-Santos for useful discussions related to our work.
We are finally grateful to Marc Sauvage for his help and for his useful and numerous comments.

\end{document}